\documentclass[12pt,reqno]{article}
\usepackage{amsmath,hyperref,amssymb}
\usepackage{subfigure,graphicx,url}

\allowdisplaybreaks \numberwithin{equation}{section}

\DeclareSymbolFont{AMSa}{U}{msa}{m}{n}
\DeclareSymbolFont{AMSb}{U}{msb}{m}{n}
\DeclareMathSymbol{\fieldR}{\mathalpha}{AMSb}{"52}

\DeclareMathOperator{\rank}{rank}
\begin{document}
\begin{flushright} \small
 ITP--UU--10/14 \\ SPIN--10/12
\end{flushright}
\bigskip
\begin{center}
 {\large\bfseries BPS black holes in N=2 D=4 gauged supergravities}\\[5mm]
Kiril Hristov$^{*,\dag}$, Hugo Looyestijn$^*$, Stefan Vandoren$^*$ \\[3mm]
 {\small\slshape
 * Institute for Theoretical Physics \emph{and} Spinoza Institute \\
 Utrecht University, 3508 TD Utrecht, The Netherlands \\
\medskip
 \dag Faculty of Physics, Sofia University, Sofia 1164, Bulgaria\\
\medskip
 {\upshape\ttfamily K.P.Hristov, H.T.Looijestijn, S.J.G.Vandoren@uu.nl}\\[3mm]}
\end{center}
\vspace{5mm} \hrule\bigskip \centerline{\bfseries Abstract}
\medskip

We construct and analyze BPS black hole solutions in gauged $N=2,
D=4$ supergravity with charged hypermultiplets. A class of
solutions can be found through spontaneous symmetry breaking in
vacua that preserve maximal supersymmetry. The resulting black
holes do not carry any hair for the scalars. We
demonstrate this with explicit examples of both asymptotically
flat and anti-de Sitter black holes.

Next, we analyze the BPS conditions for asymptotically flat black
holes with scalar hair and spherical or axial symmetry. We find solutions only in cases when the metric
contains ripples and the vector multiplet scalars become ghost-like.
We give explicit examples that can be analyzed numerically.
Finally, we comment on a way to circumvent the ghost-problem  by introducing also fermionic hair.

\bigskip
\hrule\bigskip
\newpage
\section{Introduction}

The main aim of this paper is the search for supersymmetric
four-dimensional black holes in gauged $N = 2$ supergravities in
the presence of hypermultiplets, charged under an abelian gauge
group. In the original references on BPS black holes in $D=4, N=2$ supergravity
\cite{Ferrara:1995ih,Strominger:1996kf,Ferrara:1996dd,Behrndt:1996jn,Behrndt:1997ny}, and
subsequent literature, see e.g. \cite{Denef:2000nb,LopesCardoso:2000qm,Bates:2003vx,ortinetal}, one usually considers ungauged
hypermultiplets, which then decouple from the supersymmetry variations and equations of motion
for the vector multiplet fields. We want to explore how the story
changes when the hypers couple non-trivially to the vector
multiplets via gauge couplings and scalar potentials that are allowed within gauged $N =
2$ supergravity  \cite{de Wit,DeWit:1984px,D'Auria:1990fj,Andrianopoli:1996cm,deWit:2001bk}.
For the simpler case of  minimally gauged supergravity, where no hypermultiplets are present
but only a cosmological constant or Fayet-Iliopoulos terms, asymptotically anti-de Sitter BPS black holes can be
found. This has been discussed in the literature, starting from the early references
\cite{Romans:1991nq,Kostelecky:1995ei}, or more recently in \cite{klemm-adsBH}.
We initiate here the extension to general $D=4, N=2$ gauged supergravities, including hypermultiplets.

One of our motivations comes from understanding the microscopic
entropy of (asymptotically flat) black holes. In ungauged
supergravity, arising e.g. from Calabi-Yau compactifications, this
is relatively well understood in terms of counting states in a
weakly coupled D-brane set-up \cite{Strominger:1996sh,Maldacena:1997de}, and then
extrapolating from weak to strong string coupling. In flux
compactifications, with effective gauged
supergravity actions, this picture is expected to be modified. The
most dramatic modification is probably when the dilaton is
stabilized by the fluxes, such that one cannot extrapolate between
strong and weak string coupling.

Another motivation stems from the AdS/CFT correspondence and its applications to strongly coupled field
theories. Here, finite temperature black holes that asymptote to
anti-de Sitter space-time describe the thermal behavior of the
dual field theory. Often, like e.g. in holographic superconductors, see e.g.
\cite{hol-sup} for some reviews or \cite{gauntl-bh} for more recent work, charged scalar fields are present in this black
hole geometry, providing non-trivial scalar hair\footnote{By scalar hair, in this paper, we mean a scalar field that is zero at the horizon of the black hole, but non-zero outside of the horizon. According to this definition, the vector multiplet scalars subject to the attractor mechanism in $N=2$ ungauged supergravity, do not form black holes with scalar hair. The solutions that we discuss in section \ref{sect:hair}, however, will have hair.} that can be
computed numerically. Therefore, one is in need to find large
classes of asymptotically AdS black holes with charged scalars.
This is one of the aims of this paper. Although we mostly work in
the context of supersymmetric black holes, some of our analysis in
section 3 can be carried out for finite temperature black holes as
well.

The plan of the paper is as follows. First, in section
\ref{sect:prelim}, we give a brief summary of the known black hole
solutions in $N=2$ supergravity with neutral hypermultiplets,
making a clear distinction between the asymptotically flat and
asymptotically AdS space-times. We then explain the model with
gauged hypermultiplets we are interested in and how this fits
within the framework of $N=2$ gauged supergravity.

In section \ref{sect:trivial} we first explain how one can use a
Higgs mechanism for spontaneous gauge symmetry breaking, in order
to obtain effective $N = 2$ ungauged theories from a general
gauged $N = 2$ supergravity. We keep the discussion short since
these results follow easily from our previous paper \cite{Hristov:2009uj}.
Then we show how this method can be used to embed
already known black hole solutions into gauged supergravities
and explain the physical meaning of the new solutions. We illustrate this with
an explicit example of a static, asymptotically flat black hole with the
well-known STU model and one gauged hypermultiplet (the universal
hypermultiplet). We also give examples of AdS black holes
with charged scalars, that may have applications in the emerging
field of holographic superconductivity \cite{hol-sup,gauntl-bh}.

In section \ref{sect:ansatz} we discuss in more general terms asymptotically flat,
stationary spacetimes preserving half of the supersymmetries.  We analyze
the fermion susy variations in gauged supergravity after choosing a particular ansatz for
the Killing spinor. One finds two separate cases,
defined by $T^-_{\mu \nu} = 0$ and $P^x_\Lambda = 0$,
respectively. Whereas the former case contains only Minkowski and
AdS$_4$ solutions, the latter leads to a class of solutions
that generalize the standard black hole solutions of ungauged
supergravity. We analyze this in full detail in section
\ref{sect:BLS} and give the complete set of equations that
guarantees a half-BPS solution. We then explain how this fits to
the solutions obtained in section \ref{sect:trivial}.

Finally, in section \ref{sect:hair}, we study
asymptotically flat black holes with scalar hair. We find two
separate classes of such solutions. One is a purely bosonic solution with
scalar hair, but with the shortcoming of having ghost modes in the
theory. The other class of solutions has no ghosts but along
with scalar hair we also find fermionic hair, i.e. the fermions
are not vanishing in such a vacuum.

Some of the more technical aspects of this paper, including
explicit hypermultiplet gaugings, are presented in the appendices.

\section{Preliminaries}\label{sect:prelim}

In the first part of this section, we set our notation and briefly
review the BPS black hole solutions in four-dimensional ungauged
$N=2$ supergravity. In the second part, we present the (bosonic)
action for  $N=2$ supergravity coupled to charged hypermultiplets
with abelian gaugings, and review some of the BPS black holes that
asymptote to anti-de Sitter spacetime. For a review of $N=2$ (gauged)
supergravity we refer to \cite{Andrianopoli:1996cm}, which notation we closely
follow.

\subsection{Ungauged supergravity}\label{ung-sugra}

We start by discussing the $N = 2$ Lagrangian for ungauged supergravity coupled to (abelian) vector and hypermultiplets.
The scalar fields in both these multiplets are in this case all neutral.
The theory has an action $S = \int {\rm d}^4x \sqrt{-g} \mathcal L$, and the bosonic part of
the Lagrangian $\mathcal L$ is given by
\begin{align}\label{lagr-ungauged}
\begin{split}
\mathcal L &=\frac{1}{2}R(g)+g_{i\bar
\jmath}\partial^\mu z^i
\partial_\mu
{\bar z}^{\bar \jmath} + h_{uv} \partial^\mu q^u \partial_\mu q^v +
I_{\Lambda\Sigma}(z)F_{\mu\nu}^{\Lambda}F^{\Sigma\,\mu\nu}
+\frac{1}{2}R_{\Lambda\Sigma}(z)\epsilon^{\mu\nu\rho\sigma}
F_{\mu\nu}^{\Lambda}F^{\Sigma}_{\rho\sigma} \ .
\end{split}
\end{align}
We keep the same convention for metric signatures and field strengths
 as in \cite{Hristov:2009uj}. In particular, the spacetime metric has signature $(+---)$, and we work
 in units in which the gravitational coupling constant is set to one, $\kappa^2=1$.\footnote{We corrected the sign
in front of the Einstein-Hilbert term compared to the first version in \cite{Hristov:2009uj}.}
The $z^i\, (i = 1,...,n_V)$ are the complex scalars in the vector
multiplets, with special K\"{a}hler metric $g_{i\bar \jmath}
(z,\bar{z})$. This geometry is best described in terms of
holomorphic sections $X^\Lambda(z)$ and $F_\Lambda(z),
\Lambda=0,1,...,n_V$, such that the K\"ahler potential takes the form
\begin{equation}\label{K-pot}
{\cal K}(z,\bar z)=-\ln\Big[i({\bar X}^\Lambda(\bar
z)F_\Lambda(z)-X^\Lambda(z) {\bar F}_\Lambda(\bar z))\Big]\ .
\end{equation}
When a prepotential exists, it is given by $2F=X^\Lambda
F_\Lambda$. It should be homogeneous of second degree, and one
must have that $F_\Lambda(X)=\partial F(X)/\partial X^\Lambda$.
Our general analysis does not assume the existence of a
prepotential. The complex conjugate of the ``period-matrix''
${\cal N}_{\Lambda\Sigma}$ is defined by the matrix multiplication
\begin{equation}\label{period-matrix}
{\overline {\cal N}}_{\Lambda \Sigma}\equiv \begin{pmatrix} D_iF_\Lambda \\
{\bar F}_\Lambda\end{pmatrix} \cdot {\begin{pmatrix} D_i X^\Sigma \\
{\bar X}^\Sigma\end{pmatrix}}^{-1} \ ,
\end{equation}
with ${\cal K}_i = \partial_i {\cal K}$, $D_i X^\Lambda = (\partial_i+{\cal K}_i)X^\Lambda$, and
similarly $D_i F_\Lambda = (\partial_i+{\cal K}_i)F_\Lambda$. Their imaginary and real
parts are denoted by
\begin{equation}
R_{\Lambda\Sigma}\equiv {\rm Re}\,{\cal
N}_{\Lambda\Sigma}\ ,\qquad I_{\Lambda\Sigma}\equiv {\rm Im}\,{\cal
N}_{\Lambda\Sigma}\ .
\end{equation}
The scalars in the hypermultiplet sector parametrize a
quaternion-K\"ahler manifold, whose metric can be expressed in
terms of quaternionic vielbeine. In local coordinates $q^u;
u=1,...,4n_H$, we have
\begin{equation}
h_{uv}(q)=
\mathcal{U}^{A\alpha}_u(q)\,\mathcal{U}^{B\beta}_v(q)\,\mathbb{C}_{\alpha\beta}\,\epsilon_{AB}\
,
\end{equation}
where $\mathbb{C}_{\alpha\beta}, \alpha,\beta=1,...,2n_H$ and
$\epsilon_{AB}, A, B = 1,2$ are the antisymmetric symplectic and
$SU(2)$ metrics, respectively. The value of the Ricci-scalar
curvature of the quaternionic metric is always negative and fixed in terms of Newton's
coupling constant $\kappa$. In units in which $\kappa^2=1$, which
we will use in the remainder of this paper, we have
\begin{equation}
R(h)=-8n_H(n_H+2)\ .
\end{equation}
We will discuss more on the quaternionic geometry when we introduce the
gauging at the end of this section.
Clearly, the hypermultiplet scalars
$q^u$ do not mix with the other fields (apart from the graviton)
at the level of the equations of motion, and it is therefore
consistent to set them to a constant value.

\subsection{Black holes in asymptotically Minkowski spacetime}\label{BLSsolutions}

Asymptotically flat and stationary BPS black hole solutions of
ungauged supergravity have been a very fruitful field of research
in the last decades. In absence of vector multiplets $(n_V=0)$,
with only the graviphoton present, the supersymmetric solution is
just the well-known extremal Reissner-Nordstr\"om (RN) black hole.
This solution was later generalized to include a number of vector
multiplets \cite{Ferrara:1995ih}. The most general classification of
the BPS solutions, including multicentered black holes, was given
by Behrndt, L\"ust and Sabra \cite{Behrndt:1997ny} and we will refer to those
as BLS solutions. We will briefly list the main points of the
solutions, as they will play an important role in what follows.

To characterize the black hole solutions, we first denote the
imaginary parts of the holomorphic sections by
\begin{equation}\label{eq:sections}
    \tilde{H}^{\Lambda}\equiv i (X^{\Lambda} - \bar{X}^{\Lambda})
        \ ,  \qquad  H_{\Lambda}\equiv i (F_{\Lambda} - \bar{F}_{\Lambda}) \ .
\end{equation}
We assume stationary solutions with axial  symmetry parametrized by an angular
coordinate $\varphi$. The result of the BPS analysis is that the metric takes the form\footnote{Note
that all the results are in spherical coordinates, see
\cite{Behrndt:1997ny,ortinetal} for the coordinate independent results.}
\begin{equation}\label{eq:BLSmetric}
    {\rm d} s^2 = {\rm e}^{{\cal K}} ({\rm d} t + \omega_{\varphi} {\rm d}
    \varphi)^2 - {\rm e}^{- {\cal K}} \left( {\rm d} r^2 + r^2 {\rm d}
    \Omega_2^2 \right)\ ,
\end{equation}
where ${\cal K}$ is the K\"{a}hler potential of special geometry,
defined by
\begin{equation}\label{eq:def-kahler}
{\rm e}^{-{\cal K}} = i
\left(\bar{X}^{\Lambda} F_{\Lambda} - X^{\Lambda}
\bar{F}_{\Lambda} \right)\ .
\end{equation}
The metric components and the symplectic vector $\big( \tilde{H}^{\Lambda}, H_{\Lambda} \big)$ only depend
on the radial variable $r$ and the second angular coordinate $\theta$, and the BPS conditions imply the differential equations on $\omega_{\varphi}$
\begin{align}
\begin{split}
  \frac{1}{r^2 \sin \theta} \partial_{\theta} \omega_{\varphi} =
  H_{\Lambda}\partial_r \tilde{H}^{\Lambda} - \tilde{H}^{\Lambda}
  \partial_r H_{\Lambda}\ ,\qquad
    -\frac{1}{\sin \theta} \partial_{r} \omega_{\varphi} =
  H_{\Lambda}\partial_{\theta} \tilde{H}^{\Lambda} - \tilde{H}^{\Lambda}
  \partial_{\theta} H_{\Lambda}\ .
  \end{split}
\end{align}
From this follows the integrability condition  $H_{\Lambda} \square
\tilde{H}^{\Lambda} - \tilde{H}^{\Lambda}
  \square   H_{\Lambda} = 0$, where $\square$ is the
  3-dimensional Laplacian.

What is left to specify are the gauge field strengths
$F^{\Lambda}_{\mu \nu}$. First we define the magnetic field strengths
\begin{align}\label{eq:defg}
G_{\Lambda}{}_{\mu \nu} \equiv R_{\Lambda
    \Sigma} F^{\Sigma}_{\mu \nu} - \frac 12 I_{\Lambda \Sigma}\,
  \epsilon_{\mu \nu \gamma \delta} F^{\Sigma \gamma \delta}\ ,
\end{align}
such that the Maxwell equations and Bianchi identities take the
simple form
\begin{equation}\label{eq:blsmaxwell}
    \epsilon^{\mu \nu \rho \sigma} \partial_{\nu} G_{\Lambda}{}_{\rho \sigma} =
    0, \quad \epsilon^{\mu \nu \rho \sigma} \partial_{\nu} F^{\Lambda}_{\rho \sigma} =
    0\ ,
\end{equation}
such that $(F^\Lambda, G_\Lambda)$ transforms as a vector under electric-magnetic duality transformations.

For the full solution it is enough to specify half of the
components of $F^\Lambda$ and $G_\Lambda$, since the other half can be
found from~\eqref{eq:defg}. In spherical coordinates, the BPS equations imply the
non-vanishing components\footnote{The BPS conditions also imply $F^\Lambda_{r\theta}=G_{\Lambda r\theta}=0$
due to axial symmetry.}
\begin{equation}\label{eq:FasHtilde} F^{\Lambda}_{r \varphi} =
\frac{-r^2 \sin \theta}{2}
\partial_{\theta} \tilde{H}^{\Lambda}\ , \qquad F^{\Lambda}_{\theta \varphi}
= \frac{r^2 \sin \theta}{2} \partial_r \tilde{H}^{\Lambda}\ ,
\end{equation} and
\begin{equation}\label{eq:GasH}
G_{\Lambda}{}_{r \varphi} = \frac{-r^2 \sin \theta}{2}
\partial_{\theta} H_{\Lambda}\ , \qquad G_{\Lambda}{}_{\theta \varphi} =
\frac{r^2 \sin \theta}{2} \partial_r H_{\Lambda}\ .
\end{equation}
From \eqref{eq:blsmaxwell} it now follows that $H_{\Lambda}$ and $
\tilde{H}^{\Lambda}$ are harmonic functions. With the above
identities we can always find the vector multiplet scalars $z^i$,
given that we know explicitly how they are defined in terms of the
sections $X^{\Lambda}$ and $F_{\Lambda}$.
The integration constants of the harmonic functions specify the asymptotic behavior of the fields at
the black hole horizon(s) (the constants can seen to be the black hole electric and
magnetic charges) and at spatial infinity.

The complete proof that these are indeed all the supersymmetric
black hole solutions with abelian vector multiplets and no
cosmological constant was given in \cite{ortinetal}. Note that the
BLS solutions describe half-BPS stationary spacetimes with (only
for the multi-centered cases) or without angular momentum. The
near-horizon geometry around each center is always $AdS_2 \times S^2$ with equal
radii of the two spaces, determined by the charges of the black
hole. All solutions exhibit the so-called attractor mechanism
\cite{Ferrara:1995ih}. This means that the (vector multiplet) scalar
fields get attracted to constant values at the horizon of the
black hole that only depend on the black hole charges.
As the scalars can be arbitrary constants at
infinity we also find the so-called attractor flow, i.e. the
scalars flow from their asymptotic value to the fixed constant at
the horizon. This phenomenon seems not to be related with
supersymmetry, but rather with extremality, since attractor
mechanisms have been discovered also in non-supersymmetric (but
extremal) solutions. The full classification of non-BPS
solutions and attractors is, however, more involved and is still in
progress.

\subsection{Gauged supergravity}
We now turn to the bosonic Lagrangian for gauged $N = 2$
supergravity in presence of $n_V$ abelian vector multiplets and
$n_H$ hypermultiplets, charged under the abelian gauge group (see
e.g.~\cite{Andrianopoli:1996cm} for further explanation and notation). The
effect of the gauging is to covariantize the derivatives for the
hypermultiplet scalars\footnote{For abelian gaugings, the
covariant derivative on the vector multiplet scalars is the flat
derivative, because the sections $X^\Lambda(z)$ transform in the
adjoint representation of the gauge group. One can of course
consider non-abelian gaugings, but this would complicate our black
hole analysis in subsequent sections. We leave this as a possible
generalization for future work.}, and to add a scalar potential:
\begin{align}\label{lagr}
\begin{split}
\mathcal L&=\frac{1}{2}R(g)+g_{i\bar
\jmath}\partial^\mu z^i
\partial_\mu
{\bar z}^{\bar \jmath} + h_{uv}\nabla^\mu q^u \nabla_\mu q^v\\&+
I_{\Lambda\Sigma}F_{\mu\nu}^{\Lambda}F^{\Sigma\,\mu\nu}
+\frac{1}{2}R_{\Lambda\Sigma}\epsilon^{\mu\nu\rho\sigma}
F_{\mu\nu}^{\Lambda}F^{\Sigma}_{\rho\sigma} - g^2V (z, \bar{z},
q)\ .
\end{split}
\end{align}
The covariant derivative $\nabla_{\mu} q^u \equiv
\partial_{\mu} q^u +g {\tilde k}^u_{\Lambda} A^{\Lambda}_{\mu}$
defines the gauging of some (abelian) isometries of the
quaternionic manifold with Killing vectors ${\tilde
k}^u_{\Lambda}$ and coupling constant $g$.  The scalar potential is given in terms of the Killing
vectors and the corresponding triplet of quaternionic moment maps
$P^x_{\Lambda}$ (see e.g. \cite{Andrianopoli:1996cm} for more explanation):
\begin{equation}\label{pot2}
V= 4 h_{uv}\tilde{k}^u_\Lambda \tilde{k}^v_\Sigma{\bar L}^\Lambda
L^\Sigma + (g^{i\bar \jmath}f_i^\Lambda {\bar f}_{\bar
\jmath}^\Sigma -3{\bar L}^\Lambda L^\Sigma)P^x_\Lambda
P^x_{\Sigma}\ ,
\end{equation}
where
\begin{equation}
L^\Lambda={\rm e}^{{\cal K}/2}X^\Lambda \ ,\qquad f_i^\Lambda= {\rm e}^{{\cal K}/2}D_iX^\Lambda\ .
\end{equation}
The action is invariant under the following
supersymmetry variations (up to higher order terms in fermions):
\begin{align}
\delta_\varepsilon\lambda^{iA}&=i\partial_\mu z^i
\gamma^\mu\varepsilon^A + G_{\mu\nu}^{-i}
\gamma^{\mu\nu}\epsilon^{AB}\varepsilon_B+i gg^{i\bar \jmath}{\bar
f}^\Lambda_{\bar \jmath}P^x_\Lambda\sigma_x^{AB}\varepsilon_B\
,\label{susygluino}
\\
\delta_\varepsilon \zeta_\alpha &= i\,
\mathcal{U}^{B\beta}_u\nabla_\mu q^u \gamma^\mu \varepsilon^A
\epsilon_{AB}\mathbb{C}_{\alpha\beta} +
2g\,\mathcal{U}^A_{\alpha\,u}{\tilde k}^u_\Lambda {\bar L}^\Lambda\
\varepsilon_A \ ,\label{susy-hyperino}
\\
\delta_\varepsilon \psi_{\mu A}&=\nabla_\mu\varepsilon_A +
T^-_{\mu\nu}\gamma^\nu \epsilon_{AB}\varepsilon^B + ig S_{A B}
\gamma_\mu\varepsilon^B\ ,\label{susy-gravi}
\end{align}
where $\lambda^{iA}, \zeta_\alpha$ and $\psi_{\mu A}$ are the gauginos, hyperinos and gravitinos respectively.
We have used the gravitino field strength and mass matrix
\begin{equation}\label{mass-gravitino}
T^-_{\mu\nu}\equiv 2iF^{\Lambda\,-}_{\mu\nu}\,I_{\Lambda\Sigma} L^\Sigma \ ,\qquad
S_{AB}\equiv\frac{i}{2}(\sigma_x)_{AB}P^x_\Lambda L^\Lambda\ ,
\end{equation}
and the decomposition identity
\begin{equation} \label{graviphoton-identity}
F_{\mu\nu}^{\Lambda\,-}=i{\bar L}^\Lambda
T^-_{\mu\nu}+2f_i^\Lambda G^{i\,-}_{\mu\nu}\ .
\end{equation}

The upper index ``$-$'' denotes the anti-selfdual part of the
field strengths, and in Minkowski spacetime it is complex. The
selfdual part is then obtained by complex conjugation. More
details are given in appendix~\ref{app:A}. Details on the
supercovariant derivative $\nabla_\mu\varepsilon_A$, that appears
in the supersymmetry transformation rules of the gravitinos, are
in appendix~\ref{sect:integrability}.

The fully $N=2$ supersymmetric configurations obtained from
\eqref{susygluino}-\eqref{susy-gravi} were analyzed in \cite{Hristov:2009uj}.
Two possibilities arise, namely for zero or nonzero cosmological
constant in the vacuum. For zero cosmological constant, the
different supersymmetric spacetimes are either Minkowski or
$AdS_2\times S^2$ (or its Penrose limit, the supersymmetric
pp-wave), whereas for nonzero cosmological constant only AdS$_4$
can be fully BPS. In the former case, additional constraints arise
on the scalar fields, namely (for abelian gaugings)
\begin{equation}\label{N=2Mink}
{\tilde k}^u_\Lambda L^\Lambda=0\ ,\qquad P^x_\Lambda=0\  ,
\end{equation}
together with $F^\Lambda_{\mu\nu}=0$ (Minkowski) and ${\tilde
k}^u_\Lambda F^\Lambda_{\mu\nu}=0$ ($AdS_2\times S^2$). In the
latter case, for AdS$_4$, one has the conditions
\begin{equation}\label{N=2AdS}
{\tilde k}^u_\Lambda L^\Lambda =0 \ ,\qquad P^x_\Lambda f^\Lambda_i=0\ ,\qquad \epsilon^{xyz}P^y\overline {P^z} = 0\ ,
\end{equation}
with vanishing field strengths, $F_{\mu\nu}^\Lambda=0$, and
negative scalar curvature for AdS$_4$ spacetime, $R = -12 g^2
P^x\overline{P^x}$, where $P^x\equiv P^x_\Lambda L^\Lambda$. In
all these cases, the scalars are constant or covariantly constant.
The fully supersymmetric configurations will play an important
role in the construction of 1/2 BPS black hole solutions, since
both their near horizon and asymptotic region fall into this
class. We will discuss this in detail in the following sections.

A particular class of supergravities arises in the absence of hypermultiplets.
This situation is interesting, since it allows for a bare negative cosmological constant in the Lagrangian through
the moment maps $P^x_\Lambda$ that appear in the scalar potential. It is well-known that, for $n_H=0$ and abelian gauge groups, these moment maps
can be replaced by constants (similar to Fayet-Iliopoulos terms), giving rise to a potential
\begin{equation}\label{pot3}
V= (g^{i\bar \jmath}f_i^\Lambda {\bar f}_{\bar
\jmath}^\Sigma -3{\bar L}^\Lambda L^\Sigma)P^x_\Lambda
P^x_{\Sigma}\ ,
\end{equation}
with $P^x_\Lambda$ numerical constants. When also $n_V=0$, one can take the sections $L^\Lambda$ to
be constants as well, such that the potential is negative and given by $V=-\Lambda$, with
$\Lambda=3P^x\overline{P^x}$.

\subsection{Asymptotically AdS$_4$ black holes with $n_H = 0$}\label{AdS-BH}

The construction of BPS black holes in AdS$_4$ spacetimes is
technically more involved due to the presence of the gauged
hypermultiplets, and at present there is no complete analysis for
this case. Until now, only the case with no hypermultiplets,
$n_H=0$, but with a bare cosmological constant or a potential of
the type \eqref{pot3} has been investigated in the literature
\cite{Kostelecky:1995ei,klemm,klemm-adsBH}. Static and spherically symmetric
(non-rotating) black hole solutions preserving some supersymmetry
have been constructed, but they seem to suffer from naked
singularities~\cite{Romans:1991nq,sabra-ads}. Recent developments
however show a way to construct smooth solutions~\cite{klemm-adsBH}. On the other hand there are
proper BPS black holes when one allows for a non-zero angular
momentum \cite{Kostelecky:1995ei,Caldarelli:1998hg}. The non-BPS and
non-extremal solutions, however, do allow for proper horizons also
in the non-rotating case.

Let us illustrate some of these issues in the case of static spacetimes in gauged supergravities with no
vector multiplets, so there is only a single gauge field, the graviphoton. Here we have the
AdS generalization of the Reissner-Nordstr\"om black holes (RNAdS). More explicitly, the metric in our signature is
\begin{equation}\label{adsBH}
  {\rm d} s^2 = V {\rm d} t^2 - \frac{d r^2}{V} - r^2 ({\rm d} \theta^2 + \sin^2
\theta {\rm d} \varphi^2)\ ,
\end{equation}
with
\begin{equation}\label{eq:adsmetricfunction}V(r) = 1 -
\frac{2 M}{r} + \frac{Q^2 + P^2}{r^2} - \frac{\Lambda
r^2}{3}\ .
\end{equation}
Here, $\Lambda$ is the (negative) cosmological constant and $Q$ and $P$ are the electric and magnetic charge
respectively. The field strengths are given by
\begin{equation}\label{eq:adsgaugefields}
F^-_{t r} = \frac{1}{2 r^2} \left(Q - i P
\right), \qquad F^-_{\theta \varphi} = \frac{\sin \theta}{2} \left( P
+ i Q \right)\ .
\end{equation}
For the 1/2 BPS solution the magnetic
charge is vanishing, $P = 0$ and $M = Q$~\cite{Romans:1991nq}. Of
course, this example describes naked singularities rather than
black holes. This is because $V(r)$ has no zeroes for $\Lambda <0$, so no horizons, and therefore a
naked singularity appears at $r=0$. For a genuine AdS$_4$ black hole solution we have to break
the full supersymmetry, i.e. the mass has to be free to violate
the BPS bound. If $M$ is within a certain range, as explained in
detail in e.g. \cite{Caldarelli:1998hg}, the solution has a proper
horizon and describes a  thermal AdS$_4$ black hole. There are
some BPS generalizations of these solutions to the case of
arbitrary number of vector multiplets \cite{sabra-ads}, but the
problem of naked singularities remains. For some further references on four-dimensional AdS black holes, including
the non-extremal ones, see e.g. \cite{CCLP,Morales:2006gm}.

Interestingly, recent developments in the AdS/CFT correspondence
suggest that holographic superconductors are related to
non-extremal static black holes in the presence of a charged
scalar. Such cases will arise in $N = 2$ supergravity only when
the hypermultiplets are gauged. Thus we will be able to give some
statements about this interesting class of black holes, which we
leave for section \ref{sect:examples}. In the rest of the paper we
will mainly concentrate on the asymptotically flat BPS solutions
with gauged hypers.

\section{Black holes and spontaneous symmetry breaking}\label{sect:trivial}
In this section we explain how to obtain a class of black hole
solutions in gauged supergravity, starting from known solutions in
ungauged supergravity. The main idea is simple: In gauged
supergravity, one can give expectation values to some of the
scalars (from both the vector and hypermultiplets) such that one
breaks the gauge symmetry spontaneously in a maximally
supersymmetric $N=2$ vacua, specified by the conditions
\eqref{N=2Mink} or \eqref{N=2AdS}. Let us suppose for simplicity
that the vacuum has zero cosmological constant, the argument can
be repeated for $N=2$ preserving anti-de Sitter vacua. Due to the
Higgs mechanism some of the fields become massive, and as a
consequence of the $N=2$ preserving vacua, the gravitinos remain
massless and the heavy modes form massive $N=2$ vector multiplets.
As a second step, we can set the heavy fields to zero, and the
theory gets truncated to an ungauged $N=2$ supergravity. These
truncations are consistent due to the fact that supersymmetry is
unbroken. Black hole solutions can then be found by taking any
solution of the ungauged theory and augmenting it with the massive
fields that were set to zero. In fact, it is clear from this
procedure that one can even implement a non-BPS black hole
solution of the ungauged theory into the gauged  theory. It is
also clear that this procedure works for non-abelian gaugings, as
long as it is broken spontaneously to an abelian subgroup with
residual $N=2$ supersymmetry. But for simplicity, and to streamline
with subsequent sections, we will however only consider abelian
gaugings. What is perhaps less clear, is to see if this procedure
gives the most general black hole solutions. In other words, one
can look for other solutions in which the massive scalars are
non-trivial (i.e. with scalar hair). This is the subject of
section \ref{sect:BLS}, where we investigate the conditions for
which new BPS black holes with scalar hair exist.

Let us now illustrate the above mechanism in some more detail. We
restrict ourselves first to spontaneous symmetry breaking in
Minkowski vacua, where one has $\langle P^x_\Lambda \rangle = 0$
and $\langle {\tilde k}^u_\Lambda L^\Lambda\rangle =0$ according
to \eqref{N=2Mink}. At such a point, the resulting potential is
zero, see \eqref{pot2}, as required by a Minkowski vacuum. After
the hypermultiplet scalar fields take their vacuum expectation
value, the Lagrangian~\eqref{lagr} contains a mass-term for some
of the gauge fields, given by
\begin{equation}\label{eq:vectormass}
  \mathcal L^V_{\text {mass}} = M_{\Lambda\Sigma}
  A_\mu^\Lambda A^{\mu\Sigma}\ ,\qquad M_{\Lambda\Sigma} \equiv g^2 \langle h_{uv}  \tilde k^u_\Lambda \tilde k^v_\Sigma\rangle \ .
 \end{equation}
There is no contribution to the mass matrix for the vector fields
coming from expectation values of the vector multiplet scalars,
since the gauging was chosen to be abelian. The number of massive
vectors is then given by the rank of $M_{\Lambda\Sigma}$, and as
$h_{uv}$ is positive definite, one has $\rank(M_{\Lambda\Sigma}) =
\rank(\tilde k^u_\Lambda)$. Hence, the massive vector fields are encoded by the linear combinations
${\tilde k}^u_\Lambda A_\mu^\Lambda$.
Similarly, some of the vector and
hypermultiplet scalars acquire a mass, determined by expanding the
scalar  potential,
\begin{equation}\label{eq:scalarmass}
V= 4 h_{uv}\tilde{k}^u_\Lambda \tilde{k}^v_\Sigma{\bar L}^\Lambda
L^\Sigma + (g^{i\bar \jmath}f_i^\Lambda {\bar f}_{\bar
\jmath}^\Sigma -3{\bar L}^\Lambda L^\Sigma)P^x_\Lambda
P^x_{\Sigma}\ ,
\end{equation}
to quadratic order in the fields. Then one reads off the mass
matrix, and in general there can be off-diagonal mass terms
between vector and hypermultiplet scalars.  Massive vector
multiplets can then be formed out of a massive vector, a massive
complex scalar from the vector multiplet, and 3 hypermultiplet
scalars. The fourth hypermultiplet scalar is the Goldstone mode
that is eaten by the vector field. We will illustrate this more
explicitly in some concrete examples below.

Upon setting the massive fields to zero (or integrating them out),
one obtains a supergravity theory with only massless fields.
Because of $\langle P^x_\Lambda \rangle =0$, the mass matrix for
the gravitinos is zero as follows from \eqref{mass-gravitino}.
Therefore, the resulting theory is an ungauged supergravity theory
of the type discussed in section \ref{ung-sugra}. Black hole
solutions can then be simply copied from the results in section
\ref{BLSsolutions}. By going through the Higgs mechanism in
reverse order, one can uplift this solution easily to the gauged
theory by augmenting it with the necessary expectation values of
the scalars. It is then clear that the black hole solution is not
charged with respect to the gauge fields that acquired a mass.

The situation for spontaneous symmetry breaking in an AdS vacua is
similar. To generate a negative cosmological constant from the
potential \eqref{pot2}, we must have a $\langle P^x_\Lambda\rangle
\neq 0$ in the vacuum. The conditions for unbroken $N=2$
supersymmetry  are given in \eqref{N=2AdS}. After expanding the
fields around this vacuum, one can truncate the theory further to
a Lagrangian with a bare cosmological constant, in which one can
construct black hole solutions of the type discussed in section
\ref{AdS-BH}. We will discuss an example at the end of this
section.

\subsection{Solution generating technique}\label{sect:solutiontechnique}

We now elaborate on constructing the black hole solutions more
explicitly. As explained above, the general technique is to embed
a (BPS) solution in ungauged supergravity into a gauged
supergravity. The considerations in this subsection also apply for
the more general case of non-abelian gaugings, although we are
mainly interested here in the abelian case. First, to illustrate
the systematics of our procedure, we analyze a simpler setup in
which we embed solutions from pure supergravity into a model with
vector multiplets only. Then we extend the models to include both
hypermultiplets and vector multiplets, i.e. the most general
(electrically) gauged supergravities. We always consider solutions
with vanishing fermions, i.e. the discussion concerns only the
bosonic fields.

\subsubsection{Vector multiplets}
We start from pure $N = 2$ supergravity, i.e. only the gravity
multiplet normalized as $\mathcal L=\frac{1}{2}R(g) - \frac{1}{2}
F_{\mu\nu} F^{\mu\nu} - \Lambda$. Let us assume we have found a
solution of this Lagrangian, which we denote by
$\mathring{g}_{\mu \nu}, \mathring{F}_{\mu \nu}$. We can embed
this into a supergravity theory with only vector multiplets as
follows. If we have a theory with (gauged) vector multiplets we can
find a corresponding solution to it by satisfying
\begin{equation}\label{vectors}
    \nabla_{\mu} z^i = 0\ , \qquad G_{\mu \nu}^{i} = 0\ , \qquad k^i_{\Lambda} \bar{L}^{\Lambda} = 0\ .
\end{equation}
Note that the integrability condition following from $\nabla_{\mu}
z^i = 0$ is always satisfied given the other constraints
\footnote{Also note that we have used the Killing vectors
$k^i_{\Lambda}$ that specify a gauged isometry $\nabla_{\mu} z^i =
\partial_{\mu} z^i + gk^i_{\Lambda} A^{\Lambda}_{\mu}$ on the
vector multiplet scalar manifold. These automatically vanish if the
isometry is abelian, and therefore will not be discussed further
in this paper. The formulas here are still valid for any gauged
isometry.}. We further have the relations
\begin{equation}\label{eq:trivial_vectors}g_{\mu \nu} =
\mathring{g}_{\mu \nu}\ , \quad \sqrt{2 I_{\Lambda\Sigma} \bar{L}^{\Lambda} \bar{L}^{\Sigma}}\, T^-_{\mu
\nu} = \mathring{F}^-_{\mu \nu}\ .
\end{equation}
The last equality
is to be used for determining $T^-_{\mu \nu}$. Then we can find
the solution for our new set of gauge field strengths by
$F^{\Lambda -}_{\mu \nu} = i \bar{L}^{\Lambda} T^-_{\mu \nu}$
since we already know that $G_{\mu \nu}^{i} = 0$.

The new configuration will, by construction,  satisfy all equations of
motion of the theory and will preserve the same amount of
supersymmetry (if any) as the original one. This can be checked explicitly from the supersymmetry
transformation rules \eqref{susygluino} and \eqref{susy-gravi} combined with the
results from  our previous paper \cite{Hristov:2009uj}. Indeed \eqref{vectors} comes from imposing
the vanishing of \eqref{susygluino}, while \eqref{eq:trivial_vectors} is required by the Einstein
equations. We will give a more explicit realization of this procedure in section
\ref{sect:examples}.

\subsubsection{Hypermultiplets} Given any solution of $N = 2$
supergravity with no hypermultiplets, we can obtain a new solution
with (gauged) hypermultiplets preserving the same amount of
supersymmetry as the original one. We require the theory to remain
the same in the other sectors (vector and gravity multiplets with
solution $\mathring{g}_{\mu \nu}, \mathring{F}^{\Lambda}_{\mu
\nu}, \mathring{z}^i$) and impose some additional constraints that
have to be satisfied in addition to the already given solution. We
then simply require the fields of our new theory to be
\begin{equation}\label{eq:trivialhypers}g_{\mu \nu} =
\mathring{g}_{\mu \nu}\ , \qquad F^{\Lambda}_{\mu \nu} =
\mathring{F}^{\Lambda}_{\mu \nu}\ , \qquad z^i =
\mathring{z}^i\ ,
\end{equation} under the following restriction that
has to be solved for the hypers. Here we are left with two cases:
the original theory was either with or without Fayet-Iliopoulos (FI) terms
(cosmological constant). In absence of FI terms, a new solution
after adding hypers is given by imposing the constraints:
\begin{equation}\label{hypers no FI terms}
    \nabla_{\mu} q^u = 0\  \Rightarrow \ \tilde{k}^u_{\Lambda} F^{\Lambda}_{\mu \nu} = 0\ , \qquad P^x_{\Lambda} = 0 \ ,\qquad \tilde{k}^u_{\Lambda}
    L^{\Lambda}
    = 0\ ,
\end{equation}
while in the case of original solution with FI terms we have a
solution after adding hypers (thus no longer allowing for FI terms
but keeping $P^x_{\Lambda} L^{\Lambda}$ the same) with:
\begin{equation}\label{hypers with FI terms}
    \nabla_{\mu} q^u = 0 \Rightarrow \tilde{k}^u_{\Lambda}
    F^{\Lambda}_{\mu \nu} = 0\ , \quad P^x_{\Lambda} f^{\Lambda}_i =
    0\ ,
    \quad \epsilon^{x y z} P^y_{\Lambda} P^z_{\Sigma} L^{\Lambda}
    \bar{L}^{\Sigma} = 0\ , \quad \tilde{k}^u_{\Lambda}
    L^{\Lambda} = 0\ .
\end{equation}
The new field configuration (given it can be found from the original data)
again satisfies all equations of motion and
preserves the same amount of supersymmetry as the original
one. This is true because the susy variations of gluinos and gravitinos remain the same
as in the original solution, and also the variations for the newly
introduced hyperinos are zero.

\subsubsection{Vector and hypermultiplets} This case is just
combining the two cases above. If we start with no FI terms the
new solution will be generated by imposing equations (\ref{hypers
no FI terms}) and (\ref{vectors}). If we have a solution with a
cosmological constant we need to impose (\ref{hypers with FI
terms}) and (\ref{vectors}). Then the integrability condition
following from $\nabla_{\mu} q^u = 0$ is automatically satisfied
in both cases, using relations \eqref{eq:trivial_vectors}.

\subsection{Examples}

\subsubsection{The STU model with gauged universal
hypermultiplet}\label{sect:stu}

Here we discuss an example to illustrate explicitly the procedure outlined above.
Let us
consider an $N = 2$ theory with the universal hypermultiplet. Its quaternionic
metric and isometries are given in \ref{app:C}, and isometry 5 is chosen to be gauged. This allows for asymptotically flat black holes, since we can find solutions of \eqref{hypers no FI terms}, as we shall see below\footnote{A
suitable combination of isometries 1 and 4 would also do the job.
Note that typically in string theory isometry 5 gets broken
perturbatively while 1 and 4 remain also at quantum level. For the
present discussion it is irrelevant which one we choose since we
are not trying to directly obtain the model from string theory.}.
The quaternionic Killing vector and moment maps are given by
\begin{align}
{\tilde k}_{\Lambda} &= a_{\Lambda} \left(2 R \partial_R + u \partial_u + v \partial_v + 2 D \partial_D \right),\\
{\vec P}_{\Lambda} &= a_{\Lambda} \left\{ -\frac u {\sqrt R}, \frac v {\sqrt R},
-\frac D R \right\}\ ,
\end{align}
with $a_\Lambda$ arbitrary constants.

In the vector multiplet sector we take the so-called STU model, based on the prepotential
\begin{equation}
F = \frac{X^1 X^2 X^3}{X^0}\ ,
\end{equation}
together with $z^i = \frac{X^i}{X^0}; i=1,2,3$.  The gauge group is $U(1)^3$, but it will be broken
to $U(1)^2$ in the supersymmetric Minkowski vacua, in which we construct the black hole solution.
The conditions for a fully BPS Minkowski vacuum require
$F_{\mu \nu}^{vev} = 0, z^{i, vev} = \langle z^i \rangle = \langle
b^i\rangle+i \langle v^i \rangle, u^{vev} = v^{vev} = D^{vev} = 0,
R^{vev} = \langle R \rangle $, with arbitrary constants $\langle z^i \rangle$ and $\langle R \rangle$.
Moreover, from \eqref{N=2Mink}, the
vector multiplets scalar vevs must obey $a_{\Lambda} L^{\Lambda, vev} =
0$ (which is an equation for a the $\langle z^i \rangle$'s). Then, after expanding around this vacuum, the mass terms for the scalar fields are given by the quadratic terms in
\eqref{eq:scalarmass}. Now, if we make the definition $z \equiv
a_{\Lambda} L^{\Lambda}$, we have $z^{vev} = 0$. Expanding
the first term in \eqref{eq:scalarmass} gives the mass term for
$z$,
$$\left(4
h_{uv}\tilde{k}^u_\Lambda \tilde{k}^v_\Sigma{\bar L}^\Lambda
L^\Sigma\right)^{quadratic} = 16 z \bar{z}.$$ Expanding the second
term to quadratic order gives the mass for three of the hypers:
$$\left(g^{i\bar \jmath}f_i^\Lambda {\bar f}_{\bar
\jmath}^\Sigma P^x_\Lambda P^x_{\Sigma}\right)^{quadratic} =
\frac{a_i^2 \langle v^i \rangle^2}{\langle v^1 v^2 v^3 \rangle
\langle R\rangle} \big(u^2+v^2+\frac{D^2}{\langle R\rangle}\big)\ ,$$ while
the third term vanishes at quadratic order and does not contribute
to the mass matrix of the scalars.

Therefore two of the six vector multiplet scalars become massive
(i.e. the linear combination given by our definition for $z$),
together with three of the hypers. The fourth hyper $R$ remains
massless and is eaten up by the massive gauge field $a_{\Lambda}
A^{\Lambda}_{\mu}$ (with mass $4$ given by \ref{eq:vectormass}).
Thus we are left with an effective $N = 2$ supergravity theory of
one massive and two massless vector multiplets and no
hypermultiplets, which can be further consistently truncated to
only include the massless modes. One can then search for BPS
solutions in the remaining theory and the prescription for finding
black holes is again the one given by Behrndt, L\"{u}st and Sabra
and explained in section \ref{BLSsolutions}.

We now construct the black hole solution more explicitly, following the solution generating technique
of section (\ref{sect:solutiontechnique}). For this, we need to satisfy \eqref{eq:trivialhypers}
and \eqref{hypers no FI terms}. The condition $P^x_{\Lambda} = 0$
fixes $u = v = D =0$ and the remaining non-zero Killing vectors
are $k^R_{\Lambda} = 2 R a_{\Lambda}$. Now we have to satisfy the remaining conditions
$\tilde{k}^u_{\Lambda} X^{\Lambda} = 0$ and $\tilde{k}^u_{\Lambda}
F^{\Lambda}_{\mu \nu} = 0$. To do so, we use the BLS solution of the STU model. For simplicity we
take the static limit $\omega_m = 0$, discussed in detail in
section 4.6 of \cite{Behrndt:1997ny}. The solution is fully expressed in
terms of the harmonic functions
\begin{align}
H _0 = h_0 + \frac{q_0}{r}, \qquad \tilde{H}^i &= h^i +
\frac{p^i}{r}\ ,\ \  i=1,2,3\ ,
\end{align}
under the condition that one of them is negative definite. The
sections then read
\begin{align}
X^0 = \sqrt{- \frac{\tilde{H}^1 \tilde{H}^2 \tilde{H}^3}{4 H_0}}\ ,
\qquad X^i = - i \frac{\tilde{H}^i}{2}\ ,
\end{align}
with metric function
\begin{align}
{\rm e}^{-\mathcal K} = \sqrt{-4 H_0 \tilde{H}^1 \tilde{H}^2
  \tilde{H}^3}\ .
\end{align}
In this case $F^0_{m n} = 0$ and the $F^i_{m n}$ components (here
$m,n$ are the spatial indices) are expressed solely in terms of derivatives
of $\tilde{H}^i$. After evaluating the period matrix we
obtain $F^i_{m t} = 0$ and $F^0_{m t}$ are given in terms of
derivatives of $H_0, \tilde{H}^i$. Thus the equations
$\tilde{k}^R_{\Lambda} X^{\Lambda} = 0$ and $\tilde{k}^R_{\Lambda}
F^{\Lambda}_{\mu \nu} = 0$ lead to
\begin{align}\label{STU-cond}
a_0 = 0\ , \qquad a_i h^i = 0\ , \qquad a_i p^i = 0\ .
\end{align}
The solution is qualitatively the same
as the original one, but  the charges $p^i$ and the
asymptotic constants $h^i$ are now related by \eqref{STU-cond}. So effectively,
the number of independent scalars and vectors is decreased by one, consistent with
the results from spontaneous symmetry breaking.
The usual attractor mechanism for the remaining, massless vector multiplet scalars holds while for the
hypermultiplet scalars we know that $u = v = D = 0$ and $R$ is
fixed to an arbitrary constant everywhere in spacetime with no
boundary conditions at the horizon. In other words, the hypers are not
``attracted''.

Our construction can be generalized for non-BPS solutions as well.
In the particular case of the STU model, we can obtain a
completely analogous, non-BPS,  solution by following the
procedure described in \cite{Kallosh:2006ib}.   We flip the sign of one
of the harmonic functions in \eqref{eq:sections}  such that
\begin{align}
{\rm e}^{-\mathcal K} = \sqrt{4 H_0 \tilde{H}^1 \tilde{H}^2 \tilde{H}^3}\ .
\end{align}
This solution preserves no supersymmetry, but it is
extremal. By following our procedure above, we can embed this solution into
the gauged theory.

\subsubsection{Asymptotically AdS black holes}\label{sect:examples}
Here we give a simple but yet qualitatively very general example
of how to apply the procedure outlined above to find
asymptotically anti-de Sitter black hole solutions with gauged hypers,
starting from already known black hole solutions without hypers.
In this case we start from a solution of pure
supergravity and add abelian gauged vector multiplets and
hypermultiplets. Alternatively, one can think of it as breaking the
gauge symmetry such that all hyper- and vector multiplets become
massive, and one is left with a gravity multiplet with cosmological
constant. Here we already know the full classification of black
hole solutions, as described in section \ref{AdS-BH}.

An already worked out example in section 4.2 of \cite{Hristov:2009uj} is the
case of the gauged supergravity, arising from a consistent reduction
to four dimensions of M-theory on a Sasaki-Einstein$_7$ manifold
\cite{Gauntlett:2009zw}. The resulting low-energy effective action has a
single vector multiplet and a single hypermultiplet (the universal
hypermultiplet). The special
geometry prepotential is given by
$$F = \sqrt{X^0 (X^1)^3}\ ,$$
with $X^{\Lambda} = \{ 1, \tau^2\}$, where $\tau$ is the vector
multiplet scalar, and the isometries on the UHM are given by
\begin{align}
\tilde k_0 = 24 \partial_D - 4 v \partial_u + 4 u \partial_v\ , \qquad \tilde k_1 = 24 \partial_D\ ,
\end{align}
which is combination of isometries 1 and 4 from
appendix~\ref{app:C}. The corresponding moment maps are given by
\begin{align}
P^1_0 &= \frac{4 v}{\sqrt{R}}\ ,& P^2_0 &= \frac{4 u}{\sqrt{R}}\ ,&
P^3_0 &= 4 - \frac{12 + 4 (u^2+v^2)}{R}\ ,\\
P^1_1 &= 0\ , &  P^2_1 &= 0\ ,&  P^3_1 &= - \frac{12}{R}\ .
\end{align}

Maximally supersymmetric AdS$_4$ vacua were found in \cite{Hristov:2009uj}.
The condition~\eqref{N=2AdS}
fixes the values of the vector multiplet scalar $\tau^{vev} \equiv
(\tau_1 + i \tau_2)^{vev} = i$ and two of the four hypers $u^{vev}
= v^{vev} = 0$.  The third ungauged
hyper, which is the dilaton, is fixed to the constant non-zero value $R^{vev}
= 4$. The remaining hypermultiplet scalar is an arbitrary constant $D^{vev}
= \langle D \rangle$. All the gauge fields have vanishing
expectation values at this fully supersymmetric AdS$_4$ vacuum. If
we now expand the scalar field potential \eqref{eq:scalarmass} up
to second order in fields we obtain the following mass terms
\begin{align}
V^{quadratic} = -12 + 138(\tau_1^2 + \tau_2^2) + \frac{3}{4} R^2 + 6
R \tau_2 + 10 (u^2 + v^2)\ .
\end{align}
We can see that three of the hyperscalars and the (complex) vector
multiplet scalar acquire mass. There is also a mass term $m^2 =
36$ for the gauge field $A_0+A_1$, this field thus eats up the
remaining massless hyperscalar $D$. So we observe the formation of
a massive $N=2$ vector multiplet consisting of one massive vector
and five massive scalars, and we can consistently set all these
fields to zero. The resulting Lagrangian is that of pure $N=2$ supergravity
with a cosmological constant $\Lambda = -12$. Using the static class of
black hole solutions of \eqref{adsBH}, it is straightforward to provide a solution of the gauged
supergravity theory. All the solutions described in section \ref{AdS-BH} will
also be solutions in our considered model as they obey the
Einstein-Maxwell equations  of pure supergravity.

\section{1/2 BPS solutions}\label{sect:ansatz}

In this section we will take a more systematic approach to
studying the supersymmetric solutions of~\eqref{lagr}. We search
for a solution where the expectation values of the fermions are
zero. This implies that the supersymmetry variations of the bosons
should be zero. The vanishing of the supersymmetry variations
\eqref{susygluino}-\eqref{susy-gravi} then guarantees some amount
of conserved supersymmetry. Depending on the number of independent
components of the variation parameters $\varepsilon_A$ we will
have different amount of conserved supersymmetry. Here we will
focus on particular solutions preserving (at least) 4
supercharges, i.e. half-BPS configurations. A BPS configuration
has to further satisfy the equations of motion in order to be a
real solution of the theory, so we also impose those. The
fermionic equations of motion vanish automatically, so we are left
with the equations of motion for the graviton $g_{\mu \nu}$, the
vector fields $A^{\Lambda}_{\mu}$, and the scalars $z^i$ and
$q^u$. We will come to the relation between the BPS constraints
and the field equations in due course, but we first introduce some
more relations for the Killing spinors $\varepsilon_A$.

\subsection{Killing spinor identities}
We will make use of the approach~\cite{Gauntlett:2003fk} where one
first assumes the existence of a Killing spinor. From this spinor,
various bilinears are defined, whose properties constrain the form
of the solution to a degree where a full classification is
possible. We use this method in $D=4, N=2$, which is generalizing
the main results of~\cite{ortinetal} to include hypermultiplets in
the description. As it later turns out, we cannot completely use
this method to classify all the supersymmetric configurations, but
the method nevertheless gives useful information.

We define $\varepsilon_A$ to be a Killing spinor if it solves the
gravitino variation $\delta_\varepsilon \psi_{\mu A} = 0$, defined in
\eqref{susy-gravi}, and assume $\varepsilon_A$ to be a Killing spinor
in the remainder of this article. Such spinors anti-commute, but we
can expand them on a basis of Grassmann variables and only work with
the expansion coefficients. This leads to a commuting spinor,
which we also denote with $\varepsilon_A$, and we define\footnote{We will be brief
  on some technical points of the discussion, and refer
  to~\cite{ortinetal} for more information.}
\begin{align}
\begin{split}
\overline {\varepsilon_A} &\equiv i (\varepsilon^{A})^{\dagger} \gamma_0\ ,\\
    X &\equiv \frac 12 \epsilon^{AB} \overline {\varepsilon_A} \varepsilon_B\ ,\\
 V_{\mu}{}^A{}_B &\equiv i \overline {\varepsilon^A} \gamma_\mu
 \varepsilon_B\ ,\\
\Phi_{AB\mu\nu} &\equiv \overline{\varepsilon_A} \gamma_{\mu\nu} \varepsilon_B\ .
\end{split}
\end{align}
We now show that this implies that $V^\mu \equiv V_\mu{}^A{}_A$ is a Killing
vector. For its derivatives we find
\begin{align}
\begin{split}
  \nabla_\mu V_{\nu}{}^A{}_B &= i \delta^A{}_B (T^+_{\mu\nu} X -
  T^-_{\mu\nu} \bar X) - g_{\mu\nu} (S^{AC} \epsilon_{CB} X - S_{BC}
  \epsilon^{AC} \bar X)\\
&- i (\epsilon^{AC} T^+_\mu{}^\rho \Phi_{CB\rho\nu} +
\epsilon_{BC} T^-_{\mu}{}^\rho \Phi^{AC}{}_{\nu\rho}) - (S^{AC}
\Phi_{CB\mu\nu} + S_{BC} \Phi^{AC}{}_{\mu\nu})\ .
\end{split}
\end{align}
The second and third term are traceless, so they vanish when we
compute $\nabla_\mu V_\nu$. The other terms are antisymmetric in
$\mu\nu$, so this proves
\begin{align}
  \nabla_\mu V_\nu + \nabla_\nu V_\mu = 0\ ,
\end{align}
thus $V_\mu$ is a Killing vector. We make the decomposition $
  V^A{}_{B\mu} = \frac 12 V_\mu \delta^A{}_C + \frac 1 {\sqrt 2}
  \sigma^{xA}{}_B V^x_\mu$
and using Fierz identities one finds
\begin{align}\label{eq:decomposition}
  V_\mu{}^A{}_B V_\nu{}^B{}_A = V_\mu V_\nu - \frac 12 g_{\mu\nu} V^2\ .
\end{align}
One can show that $V_\mu V^\mu = 4|X|^2$, which shows that the
Killing vector $V_\mu$ is timelike or null. For the remainder of
this paper we restrict ourselves to a timelike Killing spinor
ansatz, defined as one that leads to a timelike Killing vector. We
make this choice, as our goal is to find stationary black hole solutions,
which always have a timelike isometry~\footnote{We furthermore assume, or
  restrict to the cases, that the stationary BPS black hole has a
  time-like Killing vector which can be written as a bilinear in the
  Killing spinor.}.  In this case, by
definition, $V_\mu V^\mu = 4 |X|^2 \neq 0$, so we can
solve~\eqref{eq:decomposition} for the metric as
\begin{align}
  g_{\mu\nu}
&=\frac 1 {4|X|^2} \left(V_\mu V_\nu - 2 V_\mu^x V_\nu^x\right)\ .
\end{align}
It follows that
\begin{align}
  V_\mu = g_{\mu\nu} V^\nu = V_\mu - \frac 1{2|X|^2} V_\mu^x (V_\nu^x
  V^\nu)\ ,
\end{align}
so $V_\mu^x V^\mu = 0$. We define a time coordinate by $V^\mu
\partial_\mu = \sqrt 2 \partial_t$, which implies $V^x_t = 0$. We
decompose $V_\mu {\rm d}x^\mu = 2 \sqrt 2 X \bar X({\rm d}t + \omega)$, where
the factor in front of ${\rm d}t$ follows from $V^2 = 4 X \bar X$ and
$\omega$ has no ${\rm d}t$ component. The metric is then given by
\begin{align}\label{eq:metric}
  {\rm ds}^2 = 2 |X|^2 ({\rm d}t + \omega)^2 - \frac 1{2|X|^2}
  \gamma_{mn}{\rm d}x^m {\rm d}x^n\ ,
\end{align}
where $|X|, \omega$ and $\gamma_{mn}$ are independent of time.

Now we are ready to make a relation between the susy variations
(\ref{susygluino}--\ref{susy-gravi}) and the equations of
motion, using an elegant and simple argument of Kallosh and
Ortin \cite{Kallosh:1993wx} that was later generalized in
\cite{ortinetal}. Assuming the existence of (any amount of)
unbroken supersymmetry, one can derive a set of equations relating
the equations of motion for the bosonic fields with derivatives of
the bosonic susy variations. For our chosen theory these read:
\begin{align}
\begin{split}
\mathcal E_\Lambda^\mu i f_i^\Lambda \gamma_{\mu}
\varepsilon^A
  \epsilon_{AB} + \mathcal E_i \varepsilon_B &= 0\ ,\\
  \mathcal E_a^\mu (-i \gamma^a \varepsilon^A) + \mathcal
  E_\Lambda^\mu \left(2 \bar L^\Lambda \varepsilon_B
    \epsilon^{AB}\right) &= 0\ ,\\
\mathcal E_u \mathcal U^{u}_{\alpha A} \varepsilon^A &= 0\ ,
\end{split}
\end{align}
where $\mathcal E$ is the equation of motion for the corresponding
field in subscript. More precisely, $\mathcal{E}_a^\mu$ is the
equation for the vielbein $e_\mu^a$ (the Einstein equations),
$\mathcal{E}_\Lambda^\mu$ corresponds to $A^{\Lambda}_\mu$ (the
Maxwell equations), $\mathcal{E}_u$ corresponds to $q^u$ and
$\mathcal{E}_i$ to $z^i$. Now, let us assume that the Maxwell
equations are satisfied, $\mathcal{E}_\Lambda^\mu = 0$. If we
multiply from the left each of the remaining terms in the three
equations by $\overline{\varepsilon^B}$ and $\overline
{\varepsilon^B} \gamma^{\nu}$ and use the fact that the Killing
spinor is timelike such that $X \neq 0$ we directly obtain that
the remaining field equations are satisfied. So, apart from the
BPS conditions, only the Maxwell equations
\begin{equation}\label{maxwell}
\epsilon^{\mu \nu \rho \sigma} \partial_{\nu} G_{\Lambda\rho\sigma} =
- gh_{u v} {\tilde k}^u_{\Lambda} \nabla^{\mu}
  q^v\ ,
\end{equation}
need to be explicitly verified.

\subsection{Killing spinor ansatz}
Contracting the gaugino variation
(\ref{susygluino}) with $\varepsilon_A$ we find the condition
\begin{align}
  0 = - 2 i \bar X \nabla_\mu z^i + 4 i G^{-i}_{\rho\mu} V^\rho - i g
  k^i_\Lambda \bar L^\Lambda V_\mu - \sqrt 2 g g^{i\bar\jmath} \bar
  f_{\bar \jmath}^\Lambda P_\Lambda^x V_\mu^x\ .
\end{align}
Using this to eliminate $\nabla_\mu z^i$ and plugging back into
$\delta \lambda^{iA} = 0$ we find\footnote{One could, as done in
  e.g.~\cite{ortinetal}, eliminate the gauge fields $G^{i-}_{\rho\mu}$ to
  obtain an equivalent relation.}
\begin{align}\label{eq:susy-gaugino-contracted}
  G^{i-}_{\rho\mu} \gamma^\mu \left(2 i V^\rho \varepsilon^A - \bar
    X \gamma^\rho \epsilon^{AB} \varepsilon_B\right) + g g^{i \bar
    \jmath} \bar f_{\bar \jmath}^\Lambda P^x_\Lambda \left(- \frac 1 {\sqrt
    2} V^x_\mu \gamma^\mu \varepsilon^A + i \bar X \sigma^{xAB}
  \varepsilon_B \right)=0\ .
\end{align}
It is here that we find an important difference with the ungauged
theories. In the latter case, $g = 0$, and the second term is
absent. Then, assuming that the gauge fields $G^{i-}_{\rho\mu}$
are non-zero, one can rewrite
equation~\eqref{eq:susy-gaugino-contracted} as
\begin{align}\label{eq:ansatz}
  \varepsilon^A + i {\rm e}^{-i\alpha} \gamma_0 \epsilon^{AB} \varepsilon_B
  = 0\ ,
\end{align}
where ${\rm e}^{i \alpha} \equiv \frac X {|X|}$. One has thus
\textit{derived} the form of the Killing spinor, which is not an
ansatz anymore.

In gauged supergravity, $g \neq 0$, so there are
various ways to solve equation~\eqref{eq:susy-gaugino-contracted}.
One could, for instance, generalize~\eqref{eq:ansatz} to
\begin{align}\label{eq:gen-ansatz}
  \varepsilon^A = b \gamma^0 \epsilon^{AB} \varepsilon_B + a^x_m
  \gamma^m \sigma^{xAB} \varepsilon_B\ .
\end{align}
Plugging this back into \eqref{eq:susy-gaugino-contracted}, one obtains BPS conditions on the fields which one can then
try to solve. While this is hard  in general, it has been done in a specific case. Namely, the ansatz used for the AdS-RN black holes in minimally gauged supergravity (with a bare cosmological constant), as analyzed by
Romans~\cite{Romans:1991nq}, fits into~\eqref{eq:gen-ansatz}, but
not in~\eqref{eq:ansatz}. In fact, we will see later that
with~\eqref{eq:ansatz} one cannot find AdS black holes.

In the remainder of this article, we will use \eqref{eq:ansatz} as a
particular ansatz, hoping to find new BPS black hole solutions that
are asymptotically flat. The reader should keep in mind that more
general Killing spinors are possible, even for asymptotically flat
black holes, and therefore our procedure will most likely not be the
most general. The search for BPS black holes that asymptote to AdS$_4$, and their Killing spinors, will be postponed for future research.

\subsection{Metric and gauge field ansatz}

We will further make the extra assumption that the solution for the spacetime metric, field strengths and
scalars, is axisymmetric, i.e. there is a well-defined axis of rotation, such that $\omega = \omega_{\varphi}
{\rm d} \varphi$ lies along the angle of rotation (we choose to call it $\varphi$) in \eqref{eq:metric}. For a
stationary axisymmetric black hole solution the symmetries constrain the metric not to depend on $t$ and
$\varphi$. These symmetries also constrain the scalars and gauge field strengths to depend only on the remaining
coordinates, which we choose to call $r$ and $\theta$. We further assume $F^{\Lambda}_{r \theta} = 0$, such that
(after also using the gauge freedom) we can set $A^{\Lambda}_r = A^{\Lambda}_{\theta} = 0$ for all $\Lambda$.

\subsection{Gaugino variation}
Plugging the ansatz~\eqref{eq:ansatz} into the gaugino variation
$\delta \lambda^{iA} = 0$ gives
\begin{equation}\label{eq:Pfiszero}
   P_\Lambda^x f_i^\Lambda = 0\ ,
\end{equation}
and
\begin{equation}\label{eq:vector-conditions}
  \left({\rm e}^{-i \alpha} \partial_\mu z^i \gamma^{\mu} \gamma^0 + G^{- i}_{\mu \nu} \gamma^{\mu \nu} \right) \varepsilon_A  =
  0\ .
\end{equation}
The latter condition can be simplified further, but we will see in
what follows that it automatically becomes simpler or gets
satisfied in certain cases, so we will come back to
\eqref{eq:vector-conditions} later. We will make use of condition
\eqref{eq:Pfiszero} when solving the gravitino integrability
conditions.

\subsection{Hyperino variation}\label{sect:hyperino-variation}
With the ansatz~\eqref{eq:ansatz}, setting the hyperino variation to zero gives the condition
\begin{align}
{\rm  e}^{-i\alpha} \nabla_\mu q^u \gamma^\mu \gamma_0 + 2 g \tilde k^u_\Lambda
  \bar L^\Lambda = 0\ .
\end{align}
Using the independence of the gamma matrices, one finds
\begin{align}\label{eq:hyper-conditions}
\begin{split}
  \nabla_r q^u &= \nabla_{\theta} q^u = 0\,,\\
 \nabla_{\varphi} q^u &= \omega_{\varphi} \nabla_t q^u\,,\\
 \nabla_t q^u  &= -  \sqrt
2 g {\tilde k}^u_\Lambda \left(X \bar L^\Lambda  + \bar X
L^\Lambda
\right) , \\
0 &= \tilde k^u_\Lambda  \left( \bar X L^\Lambda - X \bar
L^\Lambda\right) \,.
\end{split}
\end{align}
Using axial symmetry and the gauge choice for the vector fields, $A^{\Lambda}_r =
A^{\Lambda}_{\theta} = 0$, it follows
that $\nabla_r q^u =
\partial_r q^u$ and $\nabla_{\theta} q^u =
\partial_{\theta} q^u$, and these both vanish from the BPS
conditions. Furthermore, the hypers cannot depend on
$t$ and $\varphi$, because this would induce such dependence also on
the vector fields and complex scalars via the Maxwell equations
\eqref{maxwell}. Thus the hypers cannot depend on any of the
space-time coordinates, so they are constant. This will be
important when we analyze the gravitino variation.

\subsection{Gravitino variation}

The gravitino equation reads
\begin{align}
  \nabla_\mu \varepsilon_A = - {\rm e}^{-i\alpha} \left(T^-_{\mu\rho}
    \gamma^\rho \delta_A{}^C + g S_{AB} \epsilon^{BC}
    \gamma_\mu\right) \gamma_0 \varepsilon_C\ .
\end{align}
We study the integrability condition which follows from this
equation. The explicit computation is presented in appendix \ref{sect:integrability}.
The main result that we will first focus on is equation \eqref{eq:pxl=0},
\begin{align}
 T^-_{\mu\nu} P^x_\Lambda L^\Lambda &= 0\ ,
\end{align}
so that there are two separate cases: $T_{\mu\nu}^- = 0$ or
$P^x_\Lambda L^\Lambda = 0$. We will study these two cases in
different subsections.

\subsubsection{$T^-_{\mu\nu} = 0$}\label{other solutions}
In this case the integrability conditions imply that the
space-time is maximally symmetric with constant scalar curvature
$P^x_\Lambda L^\Lambda$, as further explained in appendix~\ref{sect:tzero} .
This corresponds either to Minkowski
space when $P^x_\Lambda L^\Lambda = 0$, or AdS$_4$ when the scalar
curvature is non-zero. Although there might be interesting half
BPS solutions here, they will certainly not describe black holes.

\subsubsection{$P^x_\Lambda = 0$}\label{sect:BLS}

The second case is $P^x_\Lambda L^\Lambda = 0$.
We combine this identity with $P^x_\Lambda
f^\Lambda_i = 0$ from~\eqref{eq:Pfiszero}. We now obtain
\begin{align}
P^x_\Lambda  \begin{pmatrix} \bar L^\Lambda \\ f^\Lambda_i \end{pmatrix} = 0\,.
\end{align}
The matrix between brackets on the left hand side is invertible. This follows from the properties of
special geometry, and we used it also in the characterization of the maximally supersymmetric vacua in
\cite{Hristov:2009uj}. We therefore conclude that
$P^x_\Lambda = 0$. Next, we show that in this case we have enough information
to solve the gravitino variation and give the metric
functions.

From the definition~\eqref{eq:def-covariant-epsilon} for $\nabla_\mu
\varepsilon_A$, the quaternionic $Sp(1)$ connection $\omega_{\mu A}{}^B$ vanishes, as the hypers
are constant by the arguments in
section~\ref{sect:hyperino-variation}. Combining this with
$P^x_\Lambda = 0$, we see that the gravitino
variation~\eqref{susy-gravi} is precisely the same as in a theory
without hypermultiplets and vanishing FI-terms. Thus our problem
reduces to finding the most general solution of the gravitino
variation in the ungauged theory. The answer, as proven by \cite{ortinetal}, is that this is
the well-known BLS solution \cite{Behrndt:1997ny} for stationary black holes
(or naked singularities and monopoles in certain cases). Thus we
can use the BLS solution, which in fact also solves the
gaugino variation \eqref{eq:vector-conditions}. We now only have to
impose the Maxwell equations, which are not the same as in the BLS
setup, due to the gauging of the hypermultiplets.

The sections are again described by functions $H_\Lambda$ and
$\widetilde H^\Lambda$, as in~\eqref{eq:sections}, although not all of
them are harmonic. The metric and
field strengths are given
by~\eqref{eq:BLSmetric},~\eqref{eq:FasHtilde} and~\eqref{eq:GasH}. In
terms of our original description~\eqref{eq:metric}, we have that $\gamma_{mn}$ is
three-dimensional flat space and
\begin{align}
  {\rm e}^{\mathcal K} = 2 |X|^2\ .
\end{align}
In the ungauged case the Maxwell equations have no source term and the field strengths are thus
described by harmonic functions, while now in our case they will
be more complicated. We can then directly compare to the original
BLS solution described in section \ref{BLSsolutions} and see how
the new equations of motion change it. At this point we have
chosen the phase $\alpha$ in \eqref{eq:ansatz} to vanish, just
as it does in the BLS solution. We can do this without any loss of
generality since an arbitrary phase just appears in the
intermediate results for the symplectic sections
\eqref{eq:sections}, but drops out of the physical quantities such
as the metric and the field strengths.

We repeat that the Maxwell equations are given by~\eqref{maxwell},
\begin{equation}\label{eq:newMaxwell}
    \epsilon^{\mu \nu \rho \sigma} \partial_{\nu} G_{\Lambda}{}_{\rho \sigma} = - g h_{u v} {\tilde k}^u_{\Lambda} \nabla^{\mu}
    q^v\ ,
\end{equation}
with $G_{\mu\nu}$ defined as in ~\eqref{eq:defg}. Since our Bianchi identities are unmodified, and the same as in BLS, we again solve them by taking the $\tilde{H}^{\Lambda}$'s to be harmonic
functions. The difference is in the Maxwell equations.

We plug in the identities from~\eqref{eq:hyper-conditions},~\eqref{eq:BLSmetric}
and~\eqref{eq:GasH}. The components of \eqref{eq:newMaxwell} with $\mu\neq
t$ are then automatically satisfied. The only non-trivial equation
follows from $\mu=t$, and reads
\begin{align}\label{eq:maxwellforH}
\boxed{ \bigg.  \qquad \square H_{\Lambda} = - 2 g^2 {\rm e}^{-\mathcal K} h_{u v}
\tilde{k}^u_{\Lambda} \tilde{k}^v_{\Sigma} X^{\Sigma}\ ,\qquad}
\end{align}
where $\square$ is again the three-dimensional Laplacian in flat space.
The left hand side is real, and so is the right hand side, as a consequence of the last equation in \eqref{eq:hyper-conditions} and the fact that we have chosen the phase in $X/|X|$ (see \eqref{eq:ansatz} to vanish.
In other words, $X$ is real, and therefore also ${\tilde k}^u_\Lambda X^\Lambda$ is real.

We furthermore have a consistency condition for the field
strengths. The gauge potentials appear in~\eqref{eq:hyper-conditions},
but also in~\eqref{eq:GasH}, and these should lead to the same
solution. These consistency conditions were not present in the
ungauged case, since in that case there are no restrictions on $F^{\Lambda}$ from the hyperino
variation. The constraints can be easily derived from the integrability conditions of ~\eqref{eq:hyper-conditions}, and are given by
\begin{align}\label{constraints}
\begin{split}
  \tilde{k}^u_{\Lambda} \tilde{H}^{\Lambda} &= 0\ ,\\
    \tilde{k}^u_{\Lambda} F^{\Lambda}_{r \varphi} &= -
\tilde{k}^u_{\Lambda}
    \partial_{r} \left(\omega_{\varphi} {\rm e}^{\mathcal K} X^{\Lambda} \right),\\
\tilde{k}^u_{\Lambda} F^{\Lambda}_{\theta \varphi} &= -
\tilde{k}^u_{\Lambda}
    \partial_{\theta} \left(\omega_{\varphi} {\rm e}^{\mathcal K} X^{\Lambda} \right),\\
    \tilde{k}^u_{\Lambda} F^{\Lambda}_{r t} &= - \tilde{k}^u_{\Lambda}
    \partial_{r} \left( {\rm e}^{\mathcal K} X^{\Lambda} \right),\\
    \tilde{k}^u_{\Lambda} F^{\Lambda}_{\theta t} &= - \tilde{k}^u_{\Lambda}
    \partial_{\theta} \left( {\rm e}^{\mathcal K} X^{\Lambda} \right).
\end{split}
\end{align}

The first condition can always be satisfied as it merely implies
that some of the harmonic functions $\widetilde H^\Lambda$ depend on the others (remember that the hypermultiplet scalars are constant, and therefore also the Killing vectors ${\tilde k}^u_\Lambda$). In more physical terms, this constraint
decreases the number of magnetic charges by the rank of $\tilde k^u_\Lambda$. The other
constraints have to be checked against the explicit form of the
field strengths \eqref{eq:FasHtilde} and \eqref{eq:GasH}. This
cannot be done generically and has to be checked once an explicit
model is taken.

In section
\ref{sect:trivial}, we explained how the vanishing of $\tilde{k}^u_{\Lambda}
L^{\Lambda} $ and $\tilde k^u_\Lambda A_{\mu}$ led to a BPS solution using spontaneous symmetry
breaking. We can see that also from the equations of this
section. When $\tilde{k}^u_{\Lambda}
L^{\Lambda} = 0$, the right hand side of~\eqref{eq:maxwellforH} is
zero. This equation is then solved by harmonic functions
$H_{\Lambda}$. Furthermore, as $\tilde k^u_\Lambda$ is
constant, we can move it inside the derivatives in~\eqref{constraints},
so the right hand sides are zero. The left hand
sides are zero as well, as $\tilde{k}^u_{\Lambda}
F^{\Lambda}_{\mu\nu} = 0$. Finally, the condition $\tilde{k}^u_{\Lambda}
\tilde{H}^{\Lambda} = 0$ is  satisfied as
$\tilde{k}^u_{\Lambda} L^{\Lambda}$ is already real.

\section{Solutions with scalar hair}\label{sect:hair}

In this section, we search for solutions of the above BPS conditions that
do not fall in the class described in section \ref{sect:trivial}.
They describe asymptotically flat black holes and would have non-trivial profiles for the massive vector and
scalar fields, i.e. they would be distinguishable by the scalar
hair degrees of freedom outside the black hole horizon.
Remarkably, we could not find models with pure scalar hair
solutions without the need to introduce some extra features, such
as ghost modes or non-vanishing fermions. Below, we describe two examples of solutions that lead to
at least one negative eigenvalue of the K\"{a}hler metric. We show
that if we require strictly positive definite kinetic terms in the
considered models, one cannot find scalar hair solutions, but only
the ones described in section \ref{sect:trivial}. It is of course
hard to justify these ghost solutions physically. However, there
have been cases in literature where this is not necessarily a problem, e.g. in Seiberg-Witten theory
\cite{Seiberg:1994rs,Seiberg:1994aj} one has to perform duality transformation
such that the kinetic terms remain positive definite. Whether a
similar story holds in our case remains to be seen. If such
duality transformations exist they will have to map the ghost
black hole solutions of our abelian electrically gauged
supergravity to proper black hole solutions, possibly of
magnetically gauged supergravity. However, we cannot present any
direct evidence for such a possibility.

\subsection{Ghost solutions}

Before we present our examples, we start with a general comment.
We can obtain some more information from the Einstein equations.
The trace of the Einstein equations reads
\begin{align}
  R = T^q + T^z + 4 V\ ,
\end{align}
where $R$ is the Ricci scalar, and we have defined
\begin{align}
  T^q &= - 2 h_{uv} \nabla_\mu q^u \nabla^\mu q^v\ , & T^z = -2 g_{i \bar
    \jmath} \partial_\mu z^i \partial^\mu \bar z^{\bar \jmath}\ .
\end{align}
Using the BPS conditions in~\eqref{eq:hyper-conditions}, one
quickly finds $T^q = -2V$. Furthermore, as $\partial_t z^i = 0$,
we find\footnote{Recall that our spacetime signature convention is
$(+,-,-,-)$.} $T^z \geq 0$, and $V \geq 0$ by
equations~\eqref{pot2} and the condition $P^x_\Lambda = 0$. We
therefore find
\begin{align}\label{eq:traced-einstein}
  R = T^z + 2 V \geq 0\ ,
\end{align}
as long as the metric $g_{i \bar \jmath}$ is positive definite. So
the BPS conditions forbid the Ricci scalar $R$ to become negative.
In our examples below, the metric components will show some oscillatory behavior, as a consequence
of the non-linear differential equation \eqref{eq:maxwellforH}. Therefore, their derivatives, and hence the
Ricci scalar, will oscillate between positive and negative values. This would contradict the positivity bound
\eqref{eq:traced-einstein}, unless the K\"ahler metric $g_{i \bar \jmath}$ contains regions in which it is not positive definite. We now discuss this in detail with two examples.

\subsubsection{Quadratic prepotential} We start with two simple
models, which have
only one vector multiplet. They are described by the two prepotentials
\begin{align}
F = -\frac{i}{2} \left( X^0 X^0 \pm X^1 X^1 \right).
\end{align}
These lead to the special K\"ahler metrics
\begin{align}
  g_{z \bar z} = \frac {\mp 1}  {(1 \pm z \bar z)^2}\ ,
\end{align}
where $z \equiv X^1/X^0$. With the upper sign, we therefore get a negative definite
K\"{a}hler metric and the vector multiplet scalar is a
ghost field. With the lower sign, we obtain a positive definite metric. We
couple this to the universal hypermultiplet, and gauge isometry 5 from appendix
\ref{app:C}, using $A^1_\mu$ as the gauge field. The condition $P^x_{\Lambda} = 0$ fixes $u = v = D
=0$ and the only non-vanishing component of the Killing vectors is then
$\tilde{k}^R_1 = 2 R a_1$, where $a_1$ is a constant.

From the relations~\eqref{eq:sections} it follows that $X^0 =
\frac{1}{2} (H_0 - i \tilde{H}^0)$ and $X^1 = \frac{1}{2} (\pm H_1 -
i \tilde{H}^1)$. The K\"{a}hler potential~\eqref{eq:def-kahler} is then
\begin{align}
{\rm e}^{-\mathcal K} = 2 \left( X^0 \bar{X}^0 \pm X^1 \bar{X}^1 \right).
\end{align}
As we do not use $A^0_\mu$ for the gauging, $X^0$ remains
harmonic, such that even if the solution for $X^1$ is considerably
different, we still have hope of producing a black hole by having
$X^1$ as a small perturbation of the leading term $X^0$ in the
metric function ${\rm e}^{-\mathcal K}$. For simplicity, we
restrict ourself to the spherically symmetric single-centered
case, so now our constraints \eqref{constraints} lead to
$\widetilde{H}^1 = 0$ and $\tilde{k}^u_{\Lambda} F^{\Lambda}_{r t}
= - \tilde{k}^u_{\Lambda}\partial_{r}\! \left( {\rm e}^{\mathcal
K} X^{\Lambda} \right)$. The latter eventually implies that
$\widetilde{H}^0$ is constant. Since we can absorb this constant
by rescaling $H_0$, we will set $\tilde{H}^0 = 0$. Thus we are
left with $2 X^0 = H_0 =\sqrt{2} + \frac{q_0}{r}$ ($q_0 > 0$),
where we set the constant of the harmonic function to $\sqrt{2}$
to obtain canonically normalized Minkowski space as $r \rightarrow
\infty$.

The metric is given by~\eqref{eq:BLSmetric}, where
\begin{align}
{\rm e}^{-\mathcal K} = \frac{1}{2} \left( \left(\sqrt{2} +
\frac{q_0}{r}\right)^2 \pm H_1^2 \right).
\end{align}
The only undetermined function is $H_1$, which is subject
to the only equation left to be satisfied, \eqref{eq:maxwellforH},
which in this case is given by
\begin{align}\label{eq:diff-h1}
\square H_1 = \mp {\rm e}^{-\mathcal K} H_1 = \mp \frac{1}{2} \left(
\left(\sqrt{2} + \frac{q_0}{r}\right)^2 \pm H_1^2 \right) H_1\ ,
\end{align}
after setting $g |{\tilde k}| = 1$. Besides the trivial solution $H_1 = 0$
(belonging to the class solutions from section
\ref{sect:trivial}), we could not find an analytic solution to
these equations. We can analyze the differential equation as $r
\to 0$ and $r \to \infty$. As $r \to \infty$, we require ${\rm
e}^{-\mathcal
  K} \to 1$, to obtain flat space at infinity~\footnote{Perhaps one
  can relax this requirement, and generalize this analysis to include
  BPS domain walls, which have different boundary conditions. For a
  discussion in four dimensions, see e.g.~\cite{Behrndt:2001mx}.}. Likewise, we require,
as $r \to 0$, that ${\rm e}^{-\mathcal K} \to q^2 r^{-2}$, to obtain $AdS_2
\times S^2$ at the horizon. The constant $q$ (which is not necessarily equal to $q_0$) determines the (equal)
radii of AdS$_2$ and $S^2$. If we solve~\eqref{eq:diff-h1} for large
values of $r$, we have to solve $\square H_1 = \mp H_1$; for small
values of $r$ we have to solve $\square H_1 = \mp \frac 12 q^2 r^{-2} H_1$.
\begin{itemize}
\item With the upper sign (the ghost model), we find the general solution
  \begin{align}
    H_1 &= A \frac{\cos(r)}{r} + B \frac{\sin(r)}{r}\ ,  & r &\to
    \infty\ ,\\
    H_1 &= C r^{-\frac 12 - \frac 12 \sqrt{1-4 q^2}} + D r^{-\frac 12
      + \frac 12 \sqrt{1-4q^2}}\ ,& r &\to 0\ .
  \end{align}
As long as $4 q^2 < 1$, all the asymptotics are fine.

\item With the lower sign (the non-ghost model), we find the general solution
  \begin{align}\label{eq:h1-asympt}
    H_1 &= A \frac{{\rm e}^{-r}}{r} + B \frac{{\rm e}^r}{r}\ , & r &\to
    \infty\ ,\\
\label{eq:h1-asympt2}
    H_1 &= C r^{-\frac 12 - \frac 12 \sqrt{1+4 q^2}} + D r^{-\frac 12
      + \frac 12 \sqrt{1+4q^2}}\ ,& r &\to 0 \ .
  \end{align}
When $B$ is nonzero, this violates the boundary condition that  ${\rm e}^{-\mathcal
  K} \to 1$ as $r \to \infty$, so we have to set $B=0$. Likewise, we
have to set $C=0$. We will now prove that imposing such boundary conditions implies $H_1
= 0$. To do this, we use the identity
\begin{align}
  \int_0^\infty (r H_1) \partial_r^2 (r H_1) \,{\rm d}r = -\int_0^\infty \partial_r
  (r H_1) \partial_r  (r H_1) \,{\rm d}r + (rH_1) \partial_r (r H_1)
  \Big|^{r=\infty}_{r=0}\ .
\end{align}
Using~\eqref{eq:h1-asympt} and \eqref{eq:h1-asympt2} one finds that, for $B=C=0$, the boundary term
vanishes. On the left-hand side, we use~\eqref{eq:diff-h1}, and we
obtain (using $\square H_1 = r^{-1} \partial_r^2 (r H_1)$)
\begin{align}
  \int_0^\infty H_1 {\rm e}^{-\mathcal K} H_1 \,{\rm d}r = -\int_0^\infty \partial_r
  (r H_1) \partial_r  (r H_1) \,{\rm d}r\ .
\end{align}
The left-hand side is non-negative, whereas the right-hand side is
non-positive, so this proves $H_1 = 0$. This argument can easily be repeated for solutions
with only axial symmetry.

\end{itemize}

\begin{figure}[htb]
\centering
\begin{minipage}[t]{0.9 \textwidth}
\subfigure[The function $z = H_1 / H_0$.]{
\includegraphics[width=0.5 \textwidth]{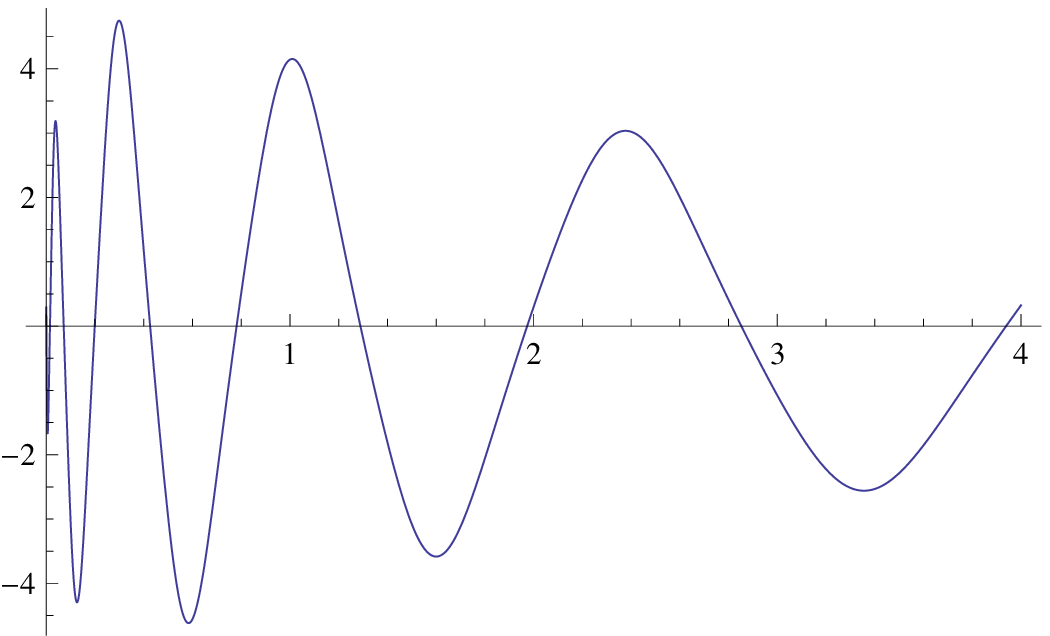}
\label{fig:subfig1} }
\subfigure[The K\"ahler potential ${\rm e}^{-\mathcal K}$.]{
\includegraphics[width=0.5 \textwidth]{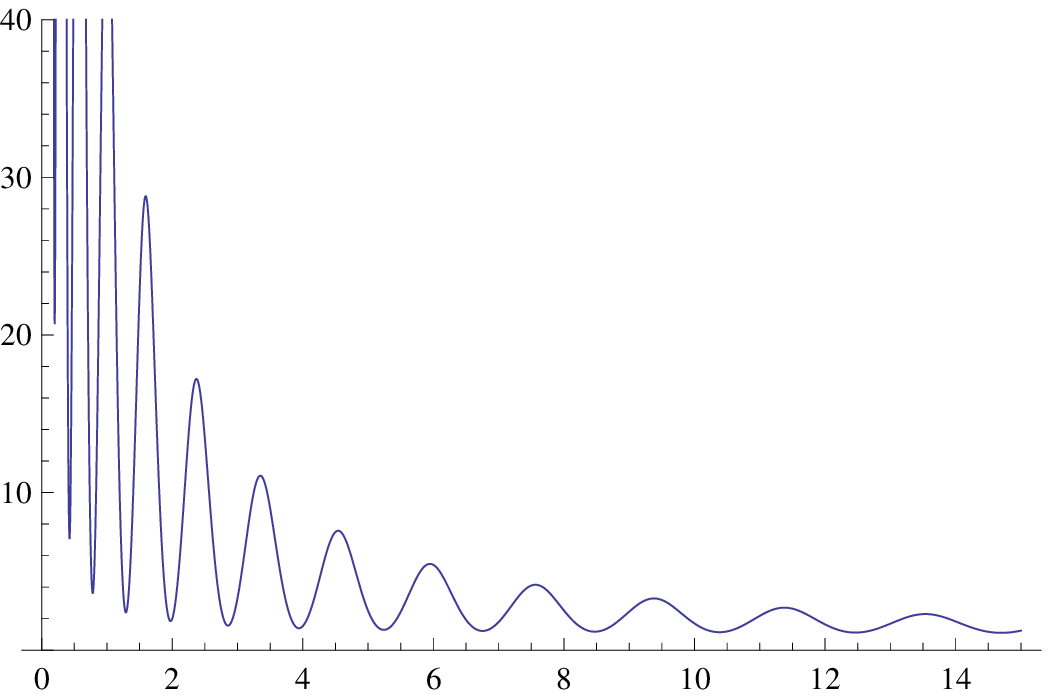}
\label{fig:subfig2}
}
\label{fig:subfigureExample} \caption{Plots of the solution to
the differential equation~\eqref{eq:diff-h1} for $q_0 =1$, using
boundary conditions $H_1(1) = 10$ and $H_1'(1) = 1$. The scalar $z$ approaches zero at the horizon at
$r=0$, and the K\"ahler potential ${\rm e}^{-\mathcal K}$ approaches $1$
as $r \to \infty$.}
\end{minipage}
\end{figure}

We can plot the solution with the upper sign numerically with
generic starting conditions, and the result is shown on
Fig.\ref{fig:subfig1}. The metric function gets
oscillatory perturbations, while having its endpoints fixed to the
desired values as shown on Fig.\ref{fig:subfig2}.

The function $H_1$ approaches zero as $r \rightarrow \infty$ in an
oscillatory fashion, which can be seen in Fig.\ref{fig:subfig1}. To investigate the behavior near the horizon
at $r=0$, we also checked that $r H_1$ approaches zero, and hence
$H_1$ diverges slower than $1/r$. Both are in agreement with the asymptotic analysis above.

The numerics further show that the metric function for negative
values of $r$ yields the expected singularity at $r = -
\frac{q_0}{\sqrt{2}}$. We conclude that this is indeed a black
hole space-time, having one electric charge $q_0$, and the
fluctuations around the usual form of the metric are due to the
effect of the abelian gauging of the hypermultiplet.

Let us now try to give a bit more physical interpretation of this
new black hole spacetime. After more careful inspection of the
solution, we see that at the horizon and asymptotically at infinity
we again have supersymmetry enhancement, since the
vector multiplet scalars are fixed to a constant value. It is
interesting that the electric charge, associated to the broken
gauge symmetry vanishes at the horizon, i.e. the black hole itself
is not charged with $q_1$ exactly as in the normal case without
ghosts. Yet there is a non-zero charge density for this charge
everywhere in the spacetime outside the black hole, which is the
qualitatively new feature of the ghost solutions. Clearly the fact
that there is non-vanishing charge density everywhere in
space-time does not change the asymptotic behavior, but it seems
that it is physically responsible for the ripples that can be
observed in the metric function on Fig. \ref{fig:subfig2} (of
course this is all related to the fact that we have propagating
ghost fields). We should note that these are not the first rippled
black hole solutions, similar behavior is found in the higher
derivative ungauged solutions, e.g. in \cite{Hubeny:2004ji}, where also
one finds ghost modes in the resulting theory. The detailed
analysis in section 4 of \cite{Hubeny:2004ji} holds in our case, i.e. the
main physical feature of the ripples is that gravitational force
changes from attractive to repulsive in some space-time points.

\subsubsection{Cubic prepotential}
The example above shows already the general qualitatively new
features of this class of black holes with ghost fields, but is
still not interesting from a string theory point of view, since
Calabi-Yau compactifications lead to cubic prepotentials of the
form
\begin{align}
F = -\frac{ \kappa_{i j k} X^i X^j X^k}{6 X^0}\ .
\end{align}
The simplest case one can consider is the STU model of section
\ref{sect:stu}. We coupled it to the universal hypermultiplet with
a single gauged isometry and found it impossible to produce any
new solutions. However, other choices of $\kappa_{i j k}$ allow
for interesting numerical solutions of \eqref{eq:maxwellforH}. For
this purpose we consider a relatively simple model with three
vector multiplets:
\begin{align}
F = \frac{(X^1)^3 - (X^1)^2 X^2 - X^1 (X^3)^2}{2 X^0}\ .
\end{align}
We again use the universal hypermultiplet and gauge the same
isometry as before, but we now use only $A^3_\mu$ for our gauging.
Again, the condition $P^x_\Lambda = 0$ fixes $u=v=D=0$, and the
only non-vanishing component of the Killing vector is $\tilde
k^R_3 =  2 Ra_3$. In parts of moduli space this model exhibits
proper Calabi-Yau behavior, i.e. the K\"{a}hler metric is positive
definite, but there are regions where $g_{i\bar \jmath}$ has
negative eigenvalues (or ${\rm e}^{-\mathcal
  K}$ becomes negative). There is no general expression for this
so-called positivity domain; one has to analyze an explicit model to
find the conditions.

For simplicity, we set $\tilde H^i = H_0 =
0$, so the non-vanishing functions are $H_i$ and $\tilde H^0$. Inverting \eqref{eq:sections} we obtain for the K\"{a}hler
potential
\begin{align}\label{eq:kahler-cy}
{\rm e}^{-\mathcal K} =  \sqrt {2 H_2} \sqrt{\tilde H^0} \left(H_1 + H_2 + \frac
  {H_3^2}{4H_2}\right)\ .
\end{align}
We see that, as is commonly encountered in these models, one has
to choose the signs of the functions $H_i$ and $\tilde H^0$ such
that this gives a real and positive quantity. With these we satisfy all conditions in \eqref{constraints} and
are left to solve \eqref{eq:maxwellforH} that explicitly reads:
\begin{align}\label{eq:diff-eq-cy}
\square H_3 = -a_3^2 \tilde H^0 \left(H_1 + H_2 + \frac {H_3^2}{4
  H_2}\right)  H_3\ ,
\end{align}
where $\tilde H^0, H_1$ and $H_2$ are harmonic functions, and we have
set $g |{\tilde k}| = 1$ for convenience.

We impose the same boundary conditions, so as $r \to \infty$, we require ${\rm
e}^{-\mathcal
  K} \to 1$, to obtain flat space at infinity. Likewise, we require,
as $r \to 0$, that ${\rm e}^{-\mathcal K} \to q^2 r^{-2}$, to obtain $AdS_2
\times S^2$ at the horizon. Using~\eqref{eq:kahler-cy}, we then find
that we have to solve
\begin{align*}
  \square H_3 &= -a_3^2 q^2 r^{-2} H_3\ ,&\text{ as }& r \to 0\ ,\\
  \square H_3 &= -a_3^2 c^2 H_3\ ,&\text{ as }& r \to \infty\ ,
\end{align*}
where $c^2$ is also a constant, specified by the asymptotics of
$\tilde H^0, H_1$ and $H_2$. We therefore again find
\begin{align}
  H_3 &= A \frac{\cos (a_3 c r)}{r} + B \frac{\sin (a_3 c
    r)}{r}\ ,&\text{ as }& r \to \infty\ .
\end{align}
These functions are oscillating; therefore the K\"ahler
potential~\eqref{eq:kahler-cy} will also oscillate. This
causes the Ricci scalar to become negative, which is in violation of
the bound~\eqref{eq:traced-einstein}. Therefore, there is always a negative eigenvalue
of the metric, corresponding to a ghost mode.

We could only find a numerical solution to this equation, and the results
are qualitatively the same as the ones on
figure~\ref{fig:subfigureExample}, so we will omit them for this
model.

It is therefore possible to find black hole solutions in these
Calabi-Yau models, but they do contain regions in which scalars become
ghost-like.

\subsection{Fermionic hair}\label{ferm-hair}

There is a different way of generating scalar hair with properly
normalized positive-definite kinetic terms. As such, we can
thereby avoid the ghost-like behavior of the previously discussed
examples. The idea is simple and works for any solution that
breaks some supersymmetry. By acting with the broken susy
generators on a bosonic solution, we will turn on the fermionic
fields to yield the fermionic zero modes. These fermionic zero
modes solve the linearized equations of motion and produce
fermionic hair. In turn, the fermionic hair sources the equations
of motion for the bosonic field, and  in particular, the scalar
field equations will have a source term which is bilinear in the
fermions. The solution of this equation produces scalar hair and
can be found explicitly by iterating again with the broken
supersymmetries. This iteration procedure stops after a finite number of
steps and produces a new solution to the full non-linear equations
of motion. By starting with a BPS black hole solution of the type
discussed in section \ref{sect:trivial}, one therefore produces
new solutions with both fermionic and scalar hair. For a
discussion on this for black holes in ungauged supergravity, see
\cite{Aichelburg:1986wv}.

The explicit realization of this idea is fairly complicated since
it requires  to explicitly find the Killing spinors preserving
supersymmetry. This can sometimes be done also just by considering
the possible bosonic and fermionic deformations of the theory, as
done in e.g. \cite{Banerjee:2009uk,Jatkar:2009yd} for black holes
in ungauged supergravity. The extension of this hair-analysis to
gauged supergravities would certainly be an interesting extension
of our work.

\section{Outlook}

In this paper, we initiated the study of BPS black holes in $N=2,D=4$ gauged supergravities.
An interesting class of solutions can be found through spontaneous symmetry breaking. They can be constructed explicitly by embedding known solutions of ungauged supergravity into the gauged theory. We also investigated
the possibility of more general BPS black hole solutions, with scalar hair. Remarkably, we could not find static solutions without ripples in the spacetime geometry and ghost-like behavior for some of the scalar fields. It would be interesting to understand this better, prove a no-go theorem or see if there are ways to circumvent the ghost-problem, e.g. along the lines of section \ref{ferm-hair}.

The BPS black hole solutions we considered in the second half of the
paper were, as a consequence of the Killing spinor ansatz
\eqref{eq:ansatz}, asymptotically flat. To find solutions which
asymptote to anti-de Sitter space, one needs to generalize the Killing spinor ansatz to for instance, \eqref{eq:gen-ansatz}. Perhaps the coupling to the hypermultiplets then allows for BPS black hole solutions in AdS$_4$ which do not contain any naked singularities.

Finally, one would like to go beyond the two-derivative approximation and study the effects of higher order curvature terms in gauged supergravities. This is interesting since the thermodynamics of the black hole, in particular the Bekenstein-Hawking law, will change. As a consequence, the microscopic interpretation within flux compactifications of string theory, might also reveal new interesting phenomena for black hole physics and quantum gravity in general.

\section*{Acknowledgments}
We thank N. Banerjee, J. de Boer, B. de Wit, S. Katmadas and A.
Sen for interesting suggestions and discussions. Furthermore, we
would like to thank J. Diederix for helpful comments on the
numerical analysis of our differential equations.

\newpage
\appendix
\section{Conventions}\label{app:A}
We mainly follow the notation and conventions from~\cite{Andrianopoli:1996cm}, so we
use a $\{+,-,-,-\}$ signature for the spacetime metric. Self-dual and anti-self-dual tensors
are defined as
\begin{align}
  F^\pm_{\mu\nu} = \frac 12 \left(F_{\mu\nu} \pm \frac i 2
    \epsilon_{\mu\nu\rho\sigma} F^{\rho\sigma}\right),
\end{align}
where $\epsilon_{0123} = 1$, and $F_{\mu\nu}\equiv \frac{1}{2}(\partial_\mu A_\nu -\partial_\nu A_\mu)$ for abelian gauge fields.

 Our gamma matrices satisfy
\begin{align}
\begin{split}
  \{\gamma_a,\gamma_b\} &= 2 \eta_{ab}\ ,\\
  [\gamma_a,\gamma_b] &\equiv 2 \gamma_{ab}\ ,\\
\gamma_5 &\equiv - i \gamma_0 \gamma_1 \gamma_2 \gamma_3 = i
\gamma^0\gamma^1 \gamma^2 \gamma^3\ .
\end{split}
\end{align}
In addition, they can be chosen such that
\begin{align}
  \gamma_0^\dagger = \gamma_0, \quad \gamma_0 \gamma_i^\dagger
  \gamma_0 = \gamma_i,\quad \gamma_5^\dagger = \gamma_5,\quad   \gamma_\mu^* = -\gamma_\mu\ ,
\end{align}
and an explicit example of such a basis is the Majorana basis, given
by
\begin{align}
  \gamma^0 &= \begin{pmatrix}0 & \sigma^2\\ \sigma^2 &
    0 \end{pmatrix},&
  \gamma^1 &= \begin{pmatrix}i \sigma^3 & 0\\ 0&i\sigma^3 \end{pmatrix},&
  \gamma^2 &= \begin{pmatrix}0 & -\sigma^2\\ \sigma^2 &
    0 \end{pmatrix},\nonumber\\
  \gamma^3 &= \begin{pmatrix}-i \sigma^1 & 0\\
    0&-i\sigma^1 \end{pmatrix},&
  \gamma_5 &= \begin{pmatrix}\sigma^2 & 0\\ 0&-\sigma^2 \end{pmatrix}\
  ,
\end{align}
where the $\sigma^i$ are the Pauli matrices.

\section{Integrability conditions}\label{sect:integrability}

The supercovariant derivative in \eqref{susy-gravi} is defined as
\begin{align}\label{eq:def-covariant-epsilon}
  \nabla_\mu \varepsilon_A = \left(\partial_\mu - \frac 14
    \omega_\mu^{ab} \gamma_{ab}\right) \varepsilon_A + \frac i 2 A_\mu
  \varepsilon_A + \omega_{\mu A}{}^B \varepsilon_B\ .
\end{align}
The connections $A_\mu$ and $\omega_{\mu A}{}^B$ are associated to
the special K\"ahler and quaternion-K\"ahler manifolds,
respectively; we refer to~\cite{Andrianopoli:1996cm} for more details. The
curvature computed from these expressions is, in a theory with
neutral vector multiplet scalars, given by~\cite{Hristov:2009uj}
\begin{align}\label{eq:comm-definitie}
\begin{split}
  [ \nabla_\mu, \nabla_\nu] \varepsilon_A &= -\frac 14 R_{\mu\nu}{}^{ab}
  \gamma_{ab} \varepsilon_A - g_{i \bar \jmath} \partial_{[\mu} z^i
  \partial_{\nu]} z^{\bar \jmath} \varepsilon_A \\&+ 2i \Omega_{uvA}{}^B \nabla_{[\mu} q^u
  \nabla_{\nu]} q^v \varepsilon_B + ig \sigma^x{}_A{}^B
  F^\Lambda_{\mu\nu} P^x_\Lambda \varepsilon_B\ .
\end{split}
\end{align}

If $\varepsilon_A$ is a Killing spinor, it obeys
  \begin{align}
  \nabla_\mu \varepsilon_A = -\epsilon_{AB} T^-_{\mu\rho} \gamma^\rho
  \varepsilon^B - i g S_{AB} \gamma_\mu \varepsilon^B\ ,
\end{align}
hence the commutator is
\begin{align}\label{eq:comm-killing}
\begin{split}
[ \nabla_\mu, \nabla_\nu] \varepsilon_A
&= -\epsilon_{AB} D_\mu T_{\nu\rho}^-
  \gamma^\rho \varepsilon^B +  \frac g 2 \sigma^x_{AB} \nabla_\mu P^x \gamma_\nu
\varepsilon^B -  (\mu\nu)\\
&+T^-_{\nu\rho} \gamma^\rho T^+_{\mu\sigma} \gamma^\sigma
\varepsilon_A - (\mu\nu) \\
&-\frac g2  T^-_{\nu \rho} \gamma^\rho \gamma_\mu
 P^x_\Lambda \bar L^\Lambda\sigma^x{}_A{}^C\varepsilon_C + \frac g 2 T^+_{\mu
  \rho} \gamma_\nu  \gamma^\rho P^x_\Lambda L^\Lambda
 \sigma^x{}_A{}^C \varepsilon_C - (\mu\nu)\\
&+\frac {g^2} 2 \left(\delta_A{}^C P^x \overline {P^x} -i\epsilon^{xyz}
  \sigma^x{}_A{}^C P^y \overline {P^z}\right) \gamma_{\mu\nu}
\varepsilon_C\ .
\end{split}
\end{align}

One can now equate~\eqref{eq:comm-killing}
to~\eqref{eq:comm-definitie}. We use~\eqref{eq:ansatz} to
eliminate $\varepsilon^A$ in terms of $\varepsilon_A$ and for
convenience define $b \equiv -i {\rm e}^{i\alpha}$ and $P^x \equiv
P^x_{\Lambda} L^{\Lambda}$. The remaining equation should hold for
any choice of $\varepsilon_A$. We can then use the independence of
the gamma matrices and the SU(2) matrices $\epsilon_{AB},
\sigma^x_{AB}$ to find the following list of conditions:
\begin{enumerate}
\item
Terms proportional to $\epsilon_{AB}$, with no gamma matrices,
\begin{align}
 b  D_\mu T^-_{\nu 0} - (\mu\nu)  = - g_{i \bar \jmath} \partial_{[\mu} z^i
  \partial_{\nu]} z^{\bar \jmath}\ .
\end{align}

\item
Terms proportional to $\epsilon_{AB}$, with two gamma matrices,
\begin{align}
b D_\mu T^-_{\nu\rho} \gamma^{\rho 0} + T^-_{\nu\rho} T^+_{\mu\sigma}
  \gamma^{\rho\sigma} - (\mu\nu)
 + \frac {g^2} 2 P^x \overline{P^x} \gamma_{\mu\nu}=-\frac 14
 R_{\mu\nu}{}^{ab} \gamma_{ab}\ .
\end{align}

\item
Terms proportional to $\sigma^x_{AB}$, no gamma matrix,
\begin{align}
\begin{split}
& \frac g 2 b  \nabla_\mu P^x g_{\nu 0}
  - (\mu\nu) + g T^-_{\mu\nu} \overline{ P^x} + g T^+_{\mu\nu} P^x\\
&=   g \left(L^\Lambda T^+_{\mu\nu} - \bar L^\Lambda T^-_{\mu\nu} - 2 i f_{\bar
  \imath}^\Lambda G^{i+}_{\mu\nu} + 2 i f_i^\Lambda
G^{i-}_{\mu\nu}\right) P^x_\Lambda\ ,
\end{split}
\end{align}
where we used that $ - \Omega_{uv}^x \nabla_{[\mu} q^u \nabla_{\nu]} q^v =
0$, which follows from~\eqref{eq:hyper-conditions}. Using $f_i^\Lambda P^x_\Lambda = 0$ from~\eqref{eq:Pfiszero} we therefore find
\begin{align}
& \frac g 2 b  \nabla_\mu P^x g_{\nu 0}
  - (\mu\nu)  =  -2g  T^-_{\mu\nu}  P^x_\Lambda \bar L^\Lambda\ .
\end{align}
We now take components $\mu=\theta$ and use $\nabla_\theta P^x = 0$
and $g_{\theta 0} = 0$. We then find $T^-_{\theta\nu} P^x = 0$,
whence $P^x = 0$ or $T^-_{\theta\nu} =0$. In the latter case also $T^-_{\mu\nu} = 0$,
because of the anti-self-duality property, and then
$T_{\mu\nu} = 0$. We conclude
\begin{align}\label{eq:pxl=0}
  T^-_{\mu\nu} P^x_\Lambda L^\Lambda = 0\ .
\end{align}

\item
Terms proportional to $\sigma^x_{AB}$, two gamma. Using~\eqref{eq:pxl=0}
we find
\begin{align}
\epsilon^{xyz} P^y \overline {P^z} \gamma_{\mu\nu}
= 0\ .
\end{align}
\end{enumerate}

To summarize: we found two cases, one with $T^-_{\mu\nu} = 0$, the other
with $P^x=0$. We now list the remaining conditions for each case.
\subsection{Case A: $T_{\mu \nu} = 0$}\label{sect:tzero}
The remaining conditions are
\begin{align}
\begin{split}
 \frac {g^2} 2 P^x \overline{P^x} \gamma_{\mu\nu}&=-\frac 14
 R_{\mu\nu}{}^{ab} \gamma_{ab}\ ,\\
g_{i\bar \jmath} \partial_{[\mu} z^i \partial_{\nu]} z^{\bar \jmath}&=
0\ ,\\
\epsilon^{xyz} P^y \overline {P^z}&=0\ .
\end{split}
\end{align}
The first condition implies that the spacetime is maximally
symmetric, with constant curvature $\propto P^x \overline {P^x}$. This
case is discussed in section~\ref{other solutions}.

\subsection{Case B: $P^x_\Lambda L^\Lambda = 0$}\label{sect:pxlzero}
The remaining conditions are
\begin{align}
\begin{split}
 b D_\mu T^-_{\nu 0} - (\mu\nu) &= - g_{i \bar \jmath} \partial_{[\mu} z^i
  \partial_{\nu]} z^{\bar \jmath}\ ,\\
b D_\mu T^-_{\nu\rho} \gamma^{\rho 0} + T^-_{\nu\rho} T^+_{\mu\sigma}
  \gamma^{\rho\sigma} - (\mu\nu) &=-\frac 14 R_{\mu\nu}{}^{ab}
  \gamma_{ab}\ .
\end{split}
\end{align}

From the second condition we find the Riemann tensor
\begin{align}
\begin{split}
R_{\mu\nu\rho\sigma} &= R^-_{\mu\nu\rho\sigma} +
R^+_{\mu\nu\rho\sigma}\ ,\\
 R^-_{\mu\nu\rho\sigma} &=  - b D_\mu T^-_{\nu \rho} e^0_\sigma +T^-_{\nu \rho} T^+_{\mu \sigma} - (\mu\nu)\nonumber\\
&- b D_\nu T^-_{\mu \sigma} e^0_\rho +T^-_{\mu \sigma} T^+_{\nu \rho} - (\mu\nu)\\
&+ b i \epsilon_{\rho\sigma}{}^{\lambda\kappa} D_\mu T_{\nu \lambda}^-
e_\kappa^0 + i
\epsilon_{\rho\sigma}{}^{\lambda\kappa} T^-_{\nu \lambda} T^+_{\mu
  \kappa} - (\mu\nu)\nonumber\ .
\end{split}
\end{align}
This
case is discussed in section~\ref{sect:BLS}.

\section{The Universal Hypermultiplet}\label{app:C}

The metric for the universal hypermultiplet is known to be
\begin{align}\label{UHM-metric}
  {\rm d}s^2 = \frac 1 {R^2} \left( {\rm d}R^2 + R \, ({\rm d}u^2 + {\rm d}v^2) + \big({\rm d}D + \frac 12 u
    {\rm d}v - \frac 12 v {\rm d}u\big)^2 \right)\ .
\end{align}
It describes the coset space $SU(2,1)/U(2)$ and therefore there are eight Killing vectors spanning the
isometry group $SU(2,1)$. In the coordinates of \eqref{UHM-metric}, they can be written as
\begin{align}
\begin{split}
  k_{a=1} &= \partial_D\ ,\\
  k_{a=2} &= \partial_u - \frac v2 \partial_D\ ,\\
  k_{a=3} &= \partial_v + \frac u2 \partial_D\ ,\\
  k_{a=4} &= -v \partial_u + u \partial_v\ ,\\
  k_{a=5} &= 2 R \partial_R + u \partial_u + v \partial_v + 2
  D \partial_D\ ,\\
  k_{a=6} &= 2 R v \partial_R + 2( u v - D) \partial_u + (-2q + v^2 -
  u^2) \partial_v + (u q + D v) \partial_D\ ,\\
  k_{a=7} &= 2 R u \partial_R + (-2q + u^2 -v^2)\partial_u + 2
  (D+uv) \partial_v + (-v q + D u) \partial_D\ ,\\
  k_{a=8} &= 2 D R \partial_R + (D u - v q) \partial_u + ( D v + u
  q) \partial_v + (D^2-q^2)\partial_D\ ,
\end{split}
\end{align}
where $q \equiv R + \frac 14 (u^2+v^2)$.

The moment maps are computed from
\begin{align}
P^x = \Omega^x_{uv} D^u k^v\ .
\end{align}
The quaternionic two-forms $\Omega^x$  satisfy $\Omega^x \Omega^y = - \frac 14 \delta^{xy} + \frac 12 \epsilon^{xyz}
  \Omega^z$, and can be written as
\renewcommand{\d}{{\rm d}}
\begin{align}
\begin{split}
  \Omega^1 &= \frac 1 {2r^{3/2}} \left( \d r \wedge \d u + \d v \wedge
    \d D + \frac v2 \d u \wedge \d v \right)\ ,\\
  \Omega^2 &= \frac 1 {2r^{3/2}} \left(- \d r \wedge \d v + \d u \wedge
    \d D + \frac u2 \d u \wedge \d v \right)\ ,\\
\Omega^3 &= \frac 1 {2r^2} \left( \d r + \d D  - \frac 12 v \d r
  \wedge \d u + \frac 12 u \d r \wedge \d v + r \d u \wedge \d v
\right)\ .
\end{split}
\end{align}
We then find the moment maps
\begin{align}
\begin{split}
  P_{a=1} &= \left\{ 0, 0, -\frac 1 {2R}\right\}, \\
  P_{a=2} &= \left\{-\frac 1 {\sqrt R},0,\frac v {2R}\right\}, \\
  P_{a=3} &= \left\{0,\frac 1 {\sqrt R},-\frac u {2R}\right\}, \\
  P_{a=4} &= \left\{\frac v{\sqrt R},\frac u {\sqrt R},1 -
   \frac{u^2+v^2}{4R}\right\},\\
  P_{a=5} &= \left\{ -\frac u {\sqrt R}, \frac v {\sqrt R}, -\frac D R
    \right\},\\
  P_{a=6} &= \left\{ \frac{2 (D-u v)}{\sqrt R}, \frac{2(q - u^2)}{
      \sqrt R},\frac{-D v - u ( 3 q -u^2-v^2)}R \right\},\\
  P_{a=7} &= \left\{ \frac{2(-q +v^2)}{
      \sqrt R}, \frac{2 (D+u v)}{\sqrt R},\frac{-D u +v ( 3 q
      -u^2-v^2)}R \right \},\\
  P_{a=8} &= \left\{\frac{v \left(-4 R+u^2+v^2\right)-4 D u}{4
      \sqrt{R}},\frac{-2 q u + u^3+ 2 D v + u v^2}{2 \sqrt R},\frac{2
      R (u^2 + v^2) -q^2 -D^2}{2R}\right\}.
\end{split}
\end{align}
These formulae are needed for some of the examples that we consider in the main text of this paper.


\begin{thebibliography}{00}

\bibitem{Ferrara:1995ih}
S.~Ferrara, R.~Kallosh and A.~Strominger,
  {\it N=2 extremal black holes},
  Phys.\ Rev.\  D {\bf 52}, 5412 (1995)
  arXiv:hep-th/9508072.

\bibitem{Strominger:1996kf}
  A.~Strominger,
  {\it Macroscopic Entropy of $N=2$ Extremal Black Holes},
  Phys.\ Lett.\  B {\bf 383} (1996) 39,
  arXiv:hep-th/9602111.

 \bibitem{Ferrara:1996dd}
  S.~Ferrara and R.~Kallosh,
  {\it Supersymmetry and Attractors},
  Phys.\ Rev.\  D {\bf 54} (1996) 1514,
  arXiv:hep-th/9602136.


\bibitem{Behrndt:1996jn}
  K.~Behrndt, G.~Lopes Cardoso, B.~de Wit, R.~Kallosh, D.~Lust and T.~Mohaupt,
  {\it Classical and quantum N=2 supersymmetric black holes},
  Nucl.\ Phys.\  B {\bf 488} (1997) 236,
  arXiv:hep-th/9610105.

\bibitem{Behrndt:1997ny}
  K.~Behrndt, D.~Lust and W.~A.~Sabra,
  {\it Stationary solutions of N = 2 supergravity},
  Nucl.\ Phys.\  B {\bf 510}, 264 (1998),
  arXiv:hep-th/9705169.


\bibitem{Denef:2000nb}
  F.~Denef,
  {\it Supergravity flows and D-brane stability},
  JHEP {\bf 0008} (2000) 050,  arXiv:hep-th/0005049.

\bibitem{LopesCardoso:2000qm}
  G.~Lopes Cardoso, B.~de Wit, J.~Carpel and T.~Mohaupt,
  {\it Stationary BPS solutions in N = 2 supergravity with R**2 interactions},
  JHEP {\bf 0012} (2000) 019,
  arXiv:hep-th/0009234.

\bibitem{Bates:2003vx}
  B.~Bates and F.~Denef,
  {\it Exact solutions for supersymmetric stationary black hole composites},
  arXiv:hep-th/0304094.

\bibitem{ortinetal}
P. Meessen and T. Ort\'{i}n, {\it The supersymmetric
configurations of N=2, d=4 supergravity coupled to vector
supermultiplets}, Nucl.\ Phys.\  B {\bf 749} (2006) 291, arXiv:hep-th/0603099;\\
M. Huebscher, P. Meessen and T. Ort\'{i}n, {\it Supersymmetric
solutions of N=2 d=4 sugra: the whole ungauged shebang}, Nucl.\
Phys.\  B {\bf 759} (2006) 228, arXiv:hep-th/0606281.

\bibitem{de Wit}
 B.~de Wit, A.~Van Proeyen,
 {\it Potentials and Symmetries of General Gauged N=2 Supergravity: Yang-Mills Models}, Nucl. Phys.\ B {\bf 245} (1984) 89;\\
  B.~de Wit, P.~G.~Lauwers, R.~Philippe, S.~Q.~Su and A.~Van Proeyen,
  {\it Gauge And Matter Fields Coupled To N=2 Supergravity},
  Phys.\ Lett.\  B {\bf 134} (1984) 37;\\
  J.~P.~Derendinger, S.~Ferrara, A.~Masiero and A.~Van Proeyen,
 {\it Yang-Mills Theories Coupled To N=2 Supergravity: Higgs And Superhiggs
  Effects In Anti-De Sitter Space},
  Phys.\ Lett.\  B {\bf 136} (1984) 354.

\bibitem{DeWit:1984px}
  B.~de Wit, P.~G.~Lauwers and A.~Van Proeyen,
  {\it Lagrangians Of N=2 Supergravity - Matter Systems,}
  Nucl.\ Phys.\  B {\bf 255}, (1985) 569.

\bibitem{D'Auria:1990fj}
  R.~D'Auria, S.~Ferrara and P.~Fr\`{e},
  {\it Special and Quaternionic Isometries: General
Couplings in N=2 Supergravity and the Scalar Potential},
  Nucl.\ Phys.\ B {\bf 359}, (1991) 705.


\bibitem{Andrianopoli:1996cm}
L. ~Adrianopoli, M. ~Bertolini, A. ~Ceresole, R. ~D'Auria, S.
~Ferrara, P. ~Fre and T. ~Magri, {\it N=2 Supergravity and N=2
Super Yang-Mills Theory on General Scalar Manifolds},
J.\ Geom.\ Phys.\  {\bf 23}, (1997) 111
arXiv:hep-th/9605032.

\bibitem{deWit:2001bk}
  B.~de Wit, M.~Ro\v{c}ek and S.~Vandoren,
  {\it Gauging isometries on hyperKaehler cones and quaternion-Kaehler
  manifolds}, Phys.\ Lett.\  B {\bf 511} (2001) 302,
  arXiv:hep-th/0104215.

\bibitem{Romans:1991nq}
  L.~J.~Romans, {\it Supersymmetric, cold and lukewarm black holes in cosmological Einstein-Maxwell theory},
  Nucl.\ Phys.\  B {\bf 383} (1992) 395, arXiv:hep-th/9203018.

\bibitem{Kostelecky:1995ei}
A. Kostelecky and M. Perry, {\it Solitonic Black Holes in Gauged
N=2 Supergravity}, Phys. Lett. {\bf B371} (1996) 191,
arXiv:hep-th/9512222.

\bibitem{klemm-adsBH}
S.~L.~Cacciatori and D.~Klemm,
  {\it Supersymmetric AdS$_4$ black holes and attractors},
  arXiv:0911.4926 [hep-th];\\
  D.~Klemm and E.~Zorzan,
  {\it The timelike half-supersymmetric backgrounds of N=2, D=4 supergravity with
  Fayet-Iliopoulos gauging},  arXiv:1003.2974 [hep-th].



\bibitem{Strominger:1996sh}
  A.~Strominger and C.~Vafa,
  {\it Microscopic Origin of the Bekenstein-Hawking Entropy},
  Phys.\ Lett.\  B {\bf 379} (1996) 99,
  arXiv:hep-th/9601029.

\bibitem{Maldacena:1997de}
  J.~M.~Maldacena, A.~Strominger and E.~Witten,
  {\it Black hole entropy in M-theory},
  JHEP {\bf 9712}, 002 (1997),
  arXiv:hep-th/9711053.


\bibitem{hol-sup}
S.~A.~Hartnoll,
{\it Lectures on holographic methods for condensed matter physics},
Class.\ Quant.\ Grav.\  {\bf 26} (2009) 224002
arXiv:0903.3246 [hep-th];\\
G. Horowitz, {\it Introduction to Holographic Superconductors},
arXiv:1002.1722 [hep-th].

\bibitem{gauntl-bh}
  J.~P.~Gauntlett, J.~Sonner and T.~Wiseman,
  {\it Holographic superconductivity in M-Theory},
  Phys.\ Rev.\ Lett.\  {\bf 103} (2009) 151601, arXiv:0907.3796 [hep-th];\\
  J.~P.~Gauntlett, J.~Sonner and T.~Wiseman,
  {\it Quantum Criticality and Holographic Superconductors in M-theory},
  JHEP {\bf 1002} (2010) 060,  arXiv:0912.0512 [hep-th].

\bibitem{Hristov:2009uj}
  K.~Hristov, H.~Looyestijn and S.~Vandoren,
  {\it Maximally supersymmetric solutions of D=4 N=2 gauged supergravity},
  JHEP {\bf 0911}, 115 (2009)
  arXiv:0909.1743 [hep-th].

\bibitem{klemm}
S. Cacciatori, D. Klemm, D. Mansi and E. Zorzan, {\it All timelike
supersymmetric solutions of N=2, D=4 gauged supergravity coupled
to abelian vector multiplets}, JHEP {\bf 0805}, 097 (2008),
arXiv:0804.0009 [hep-th];
\\
M. H\"{u}bscher, P. Meessen, T. Ort\'{i}n and S. Vaul\`{a}, {\it
$N=2$ Einstein-Yang-Mills's BPS solutions},  JHEP {\bf 0809}, 099 (2008)
, arXiv:0806.1477 [hep-th];\\
D. Klemm and E. Zorzan, {\it All null supersymmetric solutions of
N=2, D=4 gauged supergravity coupled to abelian vector multiplets
}, Class.\ Quant.\ Grav.\  {\bf 26}, 145018 (2009), arXiv:0902.4186 [hep-th].

\bibitem{sabra-ads}
W. Sabra, {\it Anti-De Sitter BPS Black Holes in N=2 Gauged
Supergravity}, arXiv:hep-th/9903143, Phys.\ Lett.\  B {\bf 458}, 36 (1999);\\
A. Chamseddine and W. Sabra, {\it Magnetic and Dyonic Black Holes
in D=4 Gauged Supergravity}, arXiv:hep-th/0003213,
Phys.\ Lett.\  B {\bf 485} (2000) 301.

\bibitem{Caldarelli:1998hg}
M. Caldarelli and D. Klemm, {\it Supersymmetry of Anti-de Sitter
Black Holes}, Nucl.\ Phys.\  B {\bf 545}, 434 (1999),
arXiv:hep-th/9808097.

\bibitem{CCLP}
  Z.~W.~Chong, M.~Cvetic, H.~Lu and C.~N.~Pope,
  {\it Charged rotating black holes in four-dimensional gauged and ungauged
  supergravities},
  Nucl.\ Phys.\  B {\bf 717} (2005) 246,
  arXiv:hep-th/0411045.

\bibitem{Morales:2006gm}
  J.~F.~Morales and H.~Samtleben,
  {\it Entropy function and attractors for AdS black holes},
  JHEP {\bf 0610} (2006) 074,
  arXiv:hep-th/0608044.

\bibitem{Kallosh:2006ib}
R. Kallosh, N. Sivanandam and M. Soroush, {\it Exact Attractive
Non-BPS STU Black Holes}, Phys.\ Rev.\  D {\bf 74}, 065008 (2006),
arXiv:hep-th/0606263.

\bibitem{Gauntlett:2009zw}
J. Gauntlett, S. Kim, O. Varela and D. Waldram, {\it Consistent
supersymmetric truncations with massive modes}, JHEP {\bf 0904}, 102 (2009),
arXiv:0901.0676 [hep-th].


\bibitem{Gauntlett:2003fk}
  J.~P.~Gauntlett and J.~B.~Gutowski,
  {\it All supersymmetric solutions of minimal gauged supergravity in five
  dimensions},
  Phys.\ Rev.\  D {\bf 68} (2003) 105009
  [Erratum-ibid.\  D {\bf 70} (2004) 089901],
  arXiv:hep-th/0304064.

\bibitem{Kallosh:1993wx}
R. Kallosh and T. Ortin, {\it Killing Spinor Identities},
arXiv:hep-th/9306085.


\bibitem{Seiberg:1994rs}
N. Seiberg and E. Witten, {\it Monopole Condensation, And
Confinement In N=2 Supersymmetric Yang-Mills Theory},
Nucl.\ Phys.\  B {\bf 426}, 19 (1994)
  [Erratum-ibid.\  B {\bf 430}, 485 (1994)]
arXiv:hep-th/9407087.


\bibitem{Seiberg:1994aj}
  N.~Seiberg and E.~Witten,
 {\it Monopoles, Duality and Chiral Symmetry
Breaking in N=2 Supersymmetric QCD},Nucl.\ Phys.\  B {\bf 431}, 484 (1994)
, arXiv:hep-th/9408099.


\bibitem{Hubeny:2004ji}
V. Hubeny, A. Maloney and M. Rangamani, {\it String-Corrected
Black Holes},
JHEP {\bf 0505}, 035 (2005),
 arXiv:hep-th/0411272.

\bibitem{Aichelburg:1986wv}
P. C. Aichelburg and F. Embacher, {\it The exact superpartners of N=2
supergravity solitons}, Phys.\ Rev.\  D {\bf 34} (1986) 3006.

\bibitem{Banerjee:2009uk}
N. Banerjee, I. Mandal and A. Sen, {\it Black Hole Hair Removal},
JHEP {\bf 0907} (2009) 091,
arXiv:0901.0359 [hep-th].

\bibitem{Jatkar:2009yd}
  D.~P.~Jatkar, A.~Sen and Y.~K.~Srivastava,
 {\it Black Hole Hair Removal:
Non-linear Analysis},
JHEP {\bf 1002} (2010) 038,
 arXiv:0907.0593 [hep-th].


\bibitem{Behrndt:2001mx}
  K.~Behrndt, G.~Lopes Cardoso and D.~Lust,
  Nucl.\ Phys.\  B {\bf 607} (2001) 391
  [arXiv:hep-th/0102128].

\end{thebibliography}
\end{document}